\documentclass[oldversion]{aa}
\usepackage{txfonts}
\usepackage{color}
\usepackage{graphicx}
\usepackage{natbib}
\usepackage{subfigure}
\begin{document}
\title{Pulsar kicks by anisotropic neutrino emission from quark matter in strong magnetic fields}
\author{I. Sagert \and J. Schaffner-Bielich}
\institute{Institute for Theoretical Physics/Astrophysics, University of Frankfurt,Max-von-Laue-Stra\ss{}e 1,
D-60438 Frankfurt, Germany}
\date{}
\abstract{We discuss a pulsar acceleration mechanism based on asymmetric neutrino emission from the direct quark Urca process in the interior of proto neutron stars. The anisotropy is caused by a strong magnetic field which polarises the spin of the electrons opposite to the field direction. Due to parity violation the neutrinos and anti-neutrinos leave the star in one direction accelerating the pulsar. We calculate for varying quark chemical potentials the kick velocity in dependence of the quark phase temperature and its radius. Ignoring neutrino quark scattering we find that within a quark phase radius of 10 km and temperatures larger than 5 MeV kick velocities of 1000km s$^{-1}$ can be reached very easily. On the other hand taking into account the small neutrino mean free paths it seems impossible to reach velocities higher than 100km s$^{-1}$ even when including effects from colour superconductivity where the neutrino quark interactions are suppressed.}
\maketitle
\section{Introduction}
Ten years after its discovery in 1967 Wyckoff and Murray \citep{Murray} estimated for the Crab pulsar a tangential velocity of 123 km s$^{-1}$ out of the centre of the supernova remnant. This high transverse speed is not very different from the value of 140km s$^{-1}$ considered today \citep{Hobbs}, nevertheless compared to other pulsars the Crab pulsar speed is still quite low. Proper motion measurements of 233 pulsars by \citet{Hobbs} imply a transverse velocity spectrum which is ranging up to $\geq 1000$km s$^{-1}$ (see also \citet{Cordes}). \citet{Brisken05} found a directly measured speed for a pulsar of v $=1083^{+103}_{-90}$km s$^{-1}$. They furthermore argued that a natal kick is required to impart the measured proper motion. The reconstruction of the initial velocity distribution of these fast pulsars is still an issue of discussion. \citet{Arzoumanian} found a two-component velocity distribution with characteristic velocities of 90km s$^{-1}$ and 500km s$^{-1}$. \citet{Bombaci} proposed that the simultaneous existence of neutron stars and quark stars could explain the bimodal distribution. On the other hand, \citet{Hobbs} find that the velocities of young pulsars are well fit by a single Maxwellian distribution with a mean velocity of 400km s$^{-1}$. Another interesting point concerning pulsar velocities is their direction. By fitting pulsar wind tori \citet{Romani} obtain nearly model independent estimates for the neutron star spin orientations which can be compared with the axes of proper motion. For the Crab and the Vela pulsar a good alignment between the velocity vector and the rotational axis is seen \citep{Brisken}. In addition, \citet{Johnston} examined the position angle of the rotational axis due to polarisation measurements of 25 pulsars and concluded that there is a strong indication for an alignment of the velocity and the spin vector.\\
Since the discovery of the high neutron star space velocities there has been a large number of suggestions to explain their origin. The general assumption is that during a certain time period in their evolution neutron stars must have experienced an accelerating kick, a so-called pulsar kick. An overview of the most common kick mechanisms can be found in e.g. \citep{Wang,Lai}. The hydrodynamical mechanism  \citep{Burrows05,Janka07} is based on an asymmetric supernova explosion where the explosion is stronger in one direction and the neutron star receives a recoil. Kick velocities of more than 1000km s$^{-1}$ are possible for successful supernova explosions \citep{Janka}. \citet{Gott} on the other hand considered the disruption of a binary system due to the explosion of one partner where the remaining star leaves with nearly its orbital velocity.  Another possible acceleration source can be asymmetric low-frequency electromagnetic radiation due to an an off-centred rotating dipole (electromagnetically driven kicks \citep{Harrison}). The combination of a rapidly rotating collapsing iron core and a magnetic field can lead to the formation of two jets at the poles. In the case of an asymmetry between these jets the supernova explosion can lead to a large kick \citep{Oran}. Neutron star acceleration can also be based on the fragmentation of a rapidly rotating core into a double proto-neutron star, where the explosion of the lighter proto-neutron star could accelerate the remaining one \citep{Colpi}.\\
A widely discussed class between acceleration mechanisms for pulsars is based on asymmetric neutrino emission \citep{Dorofeev}. Neutrinos can be a source for pulsar kicks due to the large energy release during the supernova and the following proto-neutron star evolution. A neutron star moving with a velocity of 1000 km s$^{-1}$ has a kinetic energy of $\sim 10^{49}$ erg while the energy released in neutrinos is much larger, around $\sim$10$^{53}$ erg. From momentum conservation one can see that an asymmetry of about $3\%$ could accelerate a neutron star of 1.4 M$_\odot$ to velocities of 1000 km s$^{-1}$. The momentum of neutrinos ($p_\nu$) and the neutron star ($p_{ns}$) reads:
\begin{eqnarray} 
p_\nu&=&\frac{E_\nu}{c}\simeq\frac{3\cdot 10^{53}\mbox{erg}}{3\cdot 10^{10}\mbox{cm/s}}=10^{43} \frac{\mbox{erg$\cdot$s}}{\mbox{cm}}\\
p_{ns}&=&M_{ns}\cdot v_{kick}
\nonumber\\
&=&1.4 M_\odot \cdot 10^8\frac{\mbox{cm}}{\mbox{s}}\simeq 2.8 \cdot 10^{41}\frac{\mbox{erg}\cdot\mbox{s}}{\mbox{cm}} \simeq 0.03 p_\nu.
\end{eqnarray}
Considering the violent birth of neutron stars as well as their large magnetic fields one could assume that anisotropies in the neutrino flux should be rather standard than an exception. \citet{Socrates05} discuss a radiatively driven magnetic instability which leads to a local neutrino luminosity enhancement during the Kelvin-Helmholtz cooling. \citet{Horowitz} discusses neutrino flux anisotropies stemming from parity violation in weak interactions in presence of a strong magnetic field. \citet{Horowitz_Piekarewicz}  examined amongst others anisotropies in neutrino-momentum distribution from capture of polarised electrons, whereas the resulting asymmetry of the neutrinos originates from the difference of the nucleon weak axial and vector coupling constants. The required magnetic fields were found to be in the range of $\sim 10^{16}$ Gauss throughout the whole proto neutron star. In case of smaller magnetic fields  \citet{Horowitz_Li} studied cumulative parity violation in neutrino elastic scattering from polarised neutrons in proto-neutron stars. However, it was shown by \citet{Vilenkin} that for such processes no asymmetry would be generated in thermal equilibrium even in the presence of parity violation with anisotropic scattering amplitudes (known as the "no-go theorem"). \citet{Kusenko96} as well as \citet{Goyal} argued that the tau neutrino sphere is located at a smaller radius than the one for electron neutrinos, so that electron neutrinos can oscillate to $\tau$-neutrinos in between the two spheres due to the Mikheyev - Smirnov - Wolfenstein effect (MSW-effect) and escape the proto-neutron star. The distortion of the conversion surface due to a strong magnetic field would lead to tau neutrino emission from different temperatures at different regions and therefore to a pulsar kick. A method to avoid the problem of small neutrino interaction rates completely involves sterile neutrinos. In contrast to normal neutrinos whose anisotropy is washed out for high interaction rates sterile neutrinos would keep their asymmetric distribution due to the vanishing interaction rates. Sterile neutrinos can be produced due to the MSW-effect from normal neutrinos or directly in weak processes suppressed by the square of the mixing angle (see  \citet{Kusenko_sterile, Fryer06} and references therein).\\
\citet{Ng07} recently explored the correlation between pulsar spin and proper motion assuming a single acceleration kick during the proto neutron star cooling and find that the thrust scaled proportional to the neutrino luminosity. They find that the preferred fit parameters are consistent with a magnetic field induced asymmetry of a neutrino driven kick where the characteristic timescales for the anisotropic emission are 1 - 3 s. Similar conclusions were derived by \citet{Wang} who found a kick timescale of hundreds of milliseconds to 1 s which fits well with magnetic-neutrino driven acceleration.\\
In this work we investigate such a neutrino kick mechanism for proto-neutron stars with an exotic quark matter core. Besides the traditional neutron star there are various predictions on the presence of exotic matter in the neutron star interior, as hyperons, Kaon condensation or quark matter (for an overview see e.g. \citet{Weber05}), which is expected to appear if the density inside the neutron star exceeds 2 - 3 times normal nuclear density \citep{Glendenning92}.\\
One distinguishes between so-called hybrid stars and the strange or selfbound stars. For hybrid stars, quark matter is present in a mixed phase with hadrons and can form a pure quark matter core \citep{Schertler1,Kettner00}. In the case of a strong first order phase transition between the hadronic and the quark phase, hybrid stars form a new stable solution of the Tolman-Oppenheimer-Volkoff equation \citep{Kettner00, Mishustin03, Schertler1, Fraga01,Banik03}. Besides white dwarfs and neutron stars they constitute a third family of compact stars with radii smaller than the ones of neutron stars. If strange quark matter is more stable than ordinary nuclear matter so-called strange or selfbound stars can form. They are entirely composed of absolutely stable strange quark matter covered with just a thin nucleon or strangelet crust (\citep{Alcock86, Haensel86}, for an introduction to strange stars see e.g. \citep{Schaffner-Bielich05,Weber05}. The first pure quark star was calculated by \citet{Itoh70} followed up today by a large sample of approaches for quark matter (for an overview see \citet{Schaffner-Bielich06}). The formation of quark matter can take place very early after the formation of the neutron star in the supernova explosion, during the proto-neutron star cooling stage or it can be delayed by timescales of the order of days or years depending on the mass of the metastable star ( see \citet{Drago04} and references therein). Despite their compactness quark stars can have masses up to two solar masses and therefore fit well with resent observations \citep{Ozel06} as pointed out by \citet{Alford06}.  A X-ray transient with a pulsed component of the emission having a frequency f = 1122 Hz was found recently \citet{Kaaret07}. According to \citet{Drago07} a compact star rotating with that frequency must not only contain strange matter but also be a strange star or a quark hybrid star due to r-mode instability arguments. \\
For arbitrarily high densities and temperatures quarks can be treated as a free gas of fermions. However, for neutron star interiors quark interactions due to gluon exchange become important \citep{Iwamoto81}. The consequence is that similar to electrons in superconductors, quarks can form Cooper pairs with net colour charge which is therefore referred to as colour superconductivity. The idea of colour superconducting quark matter was mentioned already around 30 years ago \citep{Barrois77,Bailin84}, was then revived simultaneously by the work of \citet{Rapp98} and \citet{Alford98} ( for overview articles about colour superconductivity see e.g. \citet{Huang,Alford01,Shovkovy04,Rischke04,Rajagopal}). Depending on the strange quark mass and the diquark coupling strengths different quark pairing patterns are preferred in dependence of the quark chemical potential and the temperature. \citet{Ruester} calculated the phase diagram for neutral quark matter with a selfconsistent treatment of quark masses. Considering a system of up, down and strange quarks where only the u and d quarks are treated as massless particles the pairing will only occur between these two flavours. The strange quark will not take part in the Cooper pairing due to its large mass and the formed colour superconducting phase is called the 2SC phase. For quark chemical potentials which are much larger than the strange quark mass one can treat all three flavours as massless. For the highest densities with quark chemical potential larger than 400 MeV the up, down and strange quarks with all colour charges will participate in the Cooper pairing and one refers to this case as the colour-flavour locked (CFL) phase \citep{Alford99,Rapp00}. A feature of the CFL phase is the absence of electrons since the charge neutrality is provided by the negatively charged strange quarks \citep{Rajagopal01}. However for non-zero temperatures electrons can appear in the CFL phase due to thermal electron-positron pair production. For temperatures larger than $T\sim$ 10 MeV the metallic CFL phase (mCFL) appears, where the electron chemical potential is non-zero \citep{Ruester04}. The so-called gapless CFL phase (gCFL) has also a non-zero density of electrons \citep{Alford04}. In this phase the pairing of strange quarks is suppressed, consequently the pairing gaps for the ds-diquark and the us-diquark condensate, respectively, vanish for specific momenta . The gCFL phase has been found to have large effects on the cooling behaviour of old neutron stars (age $\leq 10^7$ years) keeping their temperature warm for a longer time period comparing to neutron stars without the gCFL phase \citep{Jotwani}.\\
Due to the required quark chemical potentials ($\mu_q >$ 300 MeV) and low temperatures ($T <$ 100 MeV) \citep{Ruester} colour superconductivity is assumed to be realized in the interior of strange and hybrid stars. Also from the observational point of view there are arguments for its existence. The high pulsar mass of 2.10 $\pm 0.28$M$_\odot$ measured by \citet{Ozel06} indicates a very stiff equation of state which would result in too fast neutron star cooling. This fact would favour hybrid stars with colour superconducting quark matter cores as was pointed out by \citet{Klahn06}. Long gamma ray bursts with long quiescent times can be explained by a transition from hadronic matter to a 2SC phase and then from the 2SC to the CFL phase \citep{Drago06b}. \citet{Sandin07} study the effect of neutrino trapping in new-born quark stars. They find that the cores of newborn proto-neutron stars are in the 2SC state and that stable quark star solutions with CFL cores exist at low temperatures and neutrino chemical potentials. A phase transition from a hadronic phase to a quark matter phase or from ungapped to gapped quark matter provides a huge amount of energy $\epsilon \propto \Delta^2\mu_q^2$ \citep{Alford01a} which, when released asymmetrically in e.g. neutrinos, could accelerate the neutron star \citep{Lavagno07,Sagert_Pagliara}. Recently \citet{Noronha07} studied the CFL quark phase in the presence of a strong magnetic field with $eB/\mu_q2 \la 1$ (see also \citet{Fukushima07}). The authors find a smooth cross-over from the CFL to the magnetic CFL phase with oscillating pairing gaps and magnetisation for increasing magnetic fields. Furthermore \citet{Noronha07} argue that the oscillating behaviour of the magnetisation due to the Haas-van-Alphen effect may cause a large energy release during the stellar evolution. \citet{Shovkovy} discussed the asymmetry in neutrino emission by a stellar core containing spin 1 colour superconducting quark matter. Neutrinos are emitted anisotropically in space due to an asymmetric gap functions. Unfortunately, the temperature and therefore the neutrino energy for the required phase to occur is very low and leads to too small kick velocities. \citet{Blaschke} discussed neutrino emission from strange stars in the CFL phase. The mechanism is based on a beaming of neutrino emission along magnetic vortex lines and parity violation of neutrino producing weak interaction processes in magnetic fields. However, such vortices are typically not present in the CFL phase.\\
Studying asymmetric neutrino emission from the direct quark Urca process:
\begin{eqnarray}
d \longrightarrow u + e^{-} + \bar{\nu_e}\\
u + e^{-} \longrightarrow n + \nu_e
\label{dq_urca}
\end{eqnarray}
in a strong magnetic field leads also to a new problem which is not present for nucleon Urca processes. For quark matter the axial and vector coupling constants are equal and simply applying the result of \citet{Horowitz_Piekarewicz} to quark matter would give a vanishing neutrino asymmetry. A similar conclusion can be derived from the calculations by \citet{Duan}. However, the authors also find that an asymmetry is present in the differential reaction rates as soon as positrons are involved. Though there has been a lot of work done on nucleon and quark direct Urca processes in strong magnetic fields (see \citet{Dorofeev,Duan, Arras,  Lai98, Goyal, Baiko99,Riquelme05} and \citet{Iwamoto81,Steiner01,Jaikumar06,Chakrabarty, Dey98}) there are no calculations considering the angular dependence of neutrino emissivities.\\ 
In the centre of mass frame the neutrino production from electron capture by up quarks is isotropic. Just left handed particles are taking part in this process. In the presence of a strong magnetic field electrons are forced in the lowest Landau level where their spin is polarised opposite to the magnetic field direction. In the rest frame of the star the quarks have momenta in the range of 400 MeV whereas the electron momentum is less than 100 MeV. In the boosted frame the momenta of the scattered particles are beamed in the direction of the incident particle \citep{Bycling73}. Consequently, the momenta of the down quark and the neutrino are beamed in the direction of the up quark momentum creating a "neutrino emission cone". Positron capture in the centre of mass frame creates a "antineutrino cone" in the same direction. Consequently we expect a polarised neutrino emission along the magnetic axis opposite to the field direction. For simplicity we assume that the angle $\phi$ between the magnetic field axis and the rotational axis is small as was found by \citet{Ng07}.\\ 
The paper is organised as follows: In the next section we will derive analytical and numerical estimates for the degree of electron spin polarisations depending on temperature, electron chemical potential and magnetic field strength. For a sufficient acceleration the energy stored in the proto neutron star matter must equal the kinetic energy of a pulsar kick. Hence, in Sect. \ref{calc_pv} we will compute the energy density of quark matter and electrons as well as their heat capacities. Subsequently, we will calculate the kick velocities analytically and numerically for polarised electrons. In Sect. \ref{neutrino_mean} we will discuss neutrino mean free paths in hadronic matter and quark matter. Finally, in Sect. \ref{effects_csc} we comment on the acceleration mechanism for strange stars in the CFL phase and close with a summary and an outlook.
\section{Polarisation}
\label{polarisation}
\begin{figure}
\centering
\subfigure[Spin polarisation for temperature and electron chemical potential scaled by the magnetic field]{
\label{pola_general}
\includegraphics[width=0.32\textwidth,angle=270]{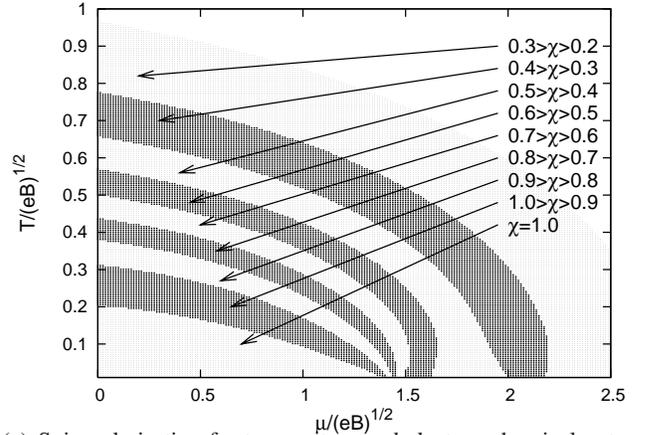} }
\subfigure[Constraint to fully polarise the electron spin for different magnetic fields in dependence of the temperature and the electron chemical potential ]{
\label{pola}
\includegraphics[width=0.32\textwidth,angle=270]{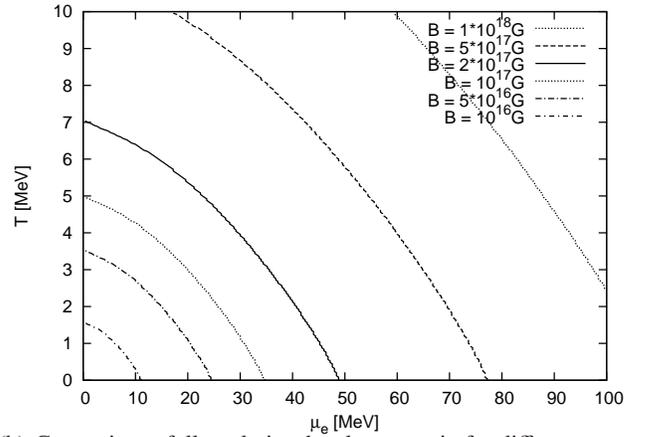}}
\caption{Critical temperature and electron chemical potential $\mu_e$ for a given magnetic field strength $B$ up to which electrons are fully spin polarised.}
\label{pol}
\end{figure}

\begin{figure}
{\centering \includegraphics[width=0.34\textwidth, angle=270]{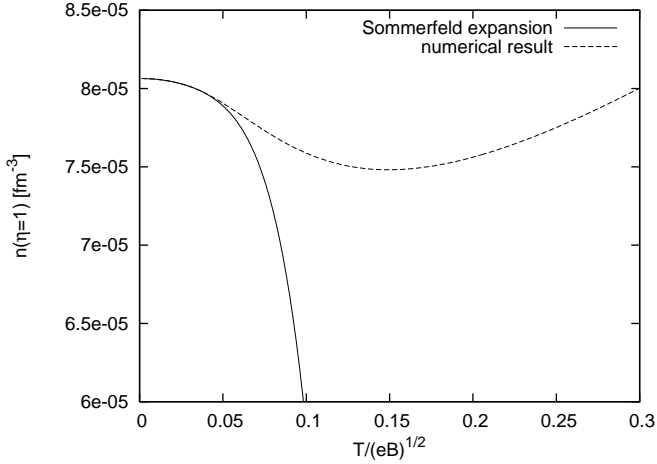} \par}
\caption{Electron number density in the first Landau level $n(\eta=1)$ in eq. (\ref{ioft}) solved numerically and analytically with a Sommerfeld expansion for $\mu=1.65 \sqrt{eB}$ in the first Landau level.}
\label{neq1}
\end{figure}

Electrons which are moving in a magnetic field larger than
\begin{eqnarray}
B_{crit}\sim\frac{m_e c^2}{e\hbar}\sim4.4\cdot10^{13}\mbox{Gauss}
\label{mag_field}
\end{eqnarray}
are situated in Landau levels perpendicular to the magnetic field axis. Their energy is quantised and can be written as \citep{Shulman01}:
\begin{eqnarray}
E^2={m_e}^2+{p_z}^2+2eB\eta={m_e}^2+{p_z}^2+eB[(2\nu+1)\mp1],
\label{energy_n}
\end{eqnarray}
where the last term in eq. (\ref{energy_n}) corresponds to the kinetic energy in the plane. The magnetic field $B$ is pointing in the positive z-direction. The Landau Level number $\eta$ is defined by its quantum number $\nu$ and the electron spin $s$ as:   
\begin{eqnarray}
\eta = \nu + \frac{1}{2} + s \mbox{ and } 
\label{landaulevel}
s & = & \left\{ \begin{array}{ccccc} +\frac{1}{2} \quad \mbox{ for  }n_+\\ -\frac{1}{2} \quad \mbox{ for } n_- \end{array}\right .
\label{spin} 
\end{eqnarray}
The electron number densities $n_{+}$ and $n_{-}$ in eq. (\ref{spin}) denote electrons with spin parallel or anti-parallel to the magnetic field direction, respectively:
\begin{eqnarray}
n_\mp&=&\frac{geB}{(2\pi)^2}\int_{\eta_{min\mp}}^{\eta_{max}}\int_0^\infty f(E)d\eta dp_z \\
&=& \frac{geB}{(2\pi)^2}\sum_{\eta_{min\mp}}^{\eta_{max}}\int_0^\infty f(E)dp_z
\label{number_den1}\\
&=&\frac{geB}{(2\pi)^2}\sum_{0}^{\nu_{max\mp}}\int_0^\infty \frac{1}{e^{(\sqrt{p^2+m_e^2+2\eta eB}-\mu_e)/T}+1} dp_z.
\label{full_dist}
\end{eqnarray}
Here $g$ is the degeneracy factor and $f(E)$ is the distribution function for fermions. In eq. (\ref{number_den1}) the integration over d$\eta$ is replaced by a summation over $\eta$ for a small number of Landau Levels. For a sufficient high magnetic field all electrons are situated in the lowest Landau Level with $\eta$=0 and spin $s=- 1/2$. In the following we will discuss the  dependence of the electron spin polarisation 
\begin{eqnarray}
\chi=\frac{n_- - n_+}{n_- + n_+}
\label{chi}
\end{eqnarray}
on temperature, electron chemical potential and magnetic field. As we will see in the next subsection only for zero temperature we have a final number of Landau levels. For $T\neq$ 0 the number of Landau levels can become very large. However, the contribution to the polarisation $\chi$ of each level decreases rapidly for large values of $\eta$ due to the small electron number density and can therefore be neglected. Figure \ref{pol} shows the numerical results for the polarisation of the electron spin for different values of the temperature $T$ and the electron chemical potential $\mu_e$ in dependence of the magnetic field strengths $B$. Figure \ref{pola_general} shows the polarisation depending on $T$ and $\mu_e$, where both are scaled by the magnetic field. It is assumed that the energy range is much larger than the electron mass so that it can be neglected. For growing temperature and chemical potential the spin polarisation decreases. Such a behaviour is expected, since the additional energy helps the electrons to overcome the energy gap of $\sqrt{2eB}$ to the next higher Landau level. However, around $T/\sqrt{eB}\sim 0.1$ and $\mu/\sqrt{eB}\sim 1.5$ the polarisation increases slightly with growing temperature. This behaviour is due to a decrease in the density of states in the first Landau level which can be calculated by 
\begin{eqnarray}
n(\eta=1)&=&\frac{2eB}{(2\pi)^2}\int_{-\infty}^{\infty} dE \frac{g(E)}{e^{(E-\mu)/T}+1}
\end{eqnarray}
where 
\begin{eqnarray}
g(E) = \left\{\begin{array}{rr}
\frac{E}{\sqrt {E^2-m_e^2-2 eB}},& E\geq \sqrt{m_e^2+2 eB}\\
0,&E < \sqrt{m_e^2+2 eB}\end{array}\right..
\end{eqnarray}
and is shown in Fig. \ref{neq1} for $\mu = 1.65\sqrt{eB}$. The minimum occurs at $T \sim 0.15\sqrt{eB}$, which it broadens and shifts to higher temperatures with growing chemical potential. The description of an electron gas in the first Landau level is analogous to the case of a one dimensional Fermion gas with a mass of $\sqrt{m_e^2+2 eB}$. As the temperatures of interest are very small in comparison to the electron chemical potential we can make a  Sommerfeld expansion of the number density:
\begin{eqnarray}
&&n(\eta=1)= \frac{2eB}{(2\pi)^2}\int_{-\infty}^{\mu} dE g(E)
\nonumber\\
&+&\frac{2eB}{(2\pi)^2}\left( 2 \sum_{n=1}^{\infty} \left(1-2^{1-2n}\right) \zeta(2n)T^{2n}\left[\frac{d^{2n-1}g(E)}{dE^{2n-1}}\right]_{E=\mu}\right)\\
&\sim&\frac{2eB}{(2\pi)^2}\left( \sqrt{\mu^2 - m_e^2 - 2eB} - \frac{1}{3}\pi^2 T^2 \frac{eB}{\left(\mu^2-m_e^2-2eB\right)^{3/2}}\right)
\nonumber\\
&-&\frac{2eB}{(2\pi)^2}\left( \frac{7}{30}\pi^4 T^4 \frac{2\mu^2 eB + (eB)^2}{\left(\mu^2-m_e^2-2eB\right)^{7/2}}\right)
\nonumber\\
&-&\frac{2eB}{(2\pi)^2}\left(\frac{31}{42}\pi^6 T^6 \frac{2\mu^4 eB+ 6(eB)^2\mu^2 + (eB)^3}{ \left(\mu^2-m_e^2-2eB\right)^{11/2}}\right).
\label{ioft}
\end{eqnarray}
The result is also plotted in Fig. \ref{neq1} and confirms the decrease in the electron density of states with temperature. Figure \ref{pola} shows the constraints on the temperature and the electron chemical potential to fully polarise the electron spin, that is $\chi = 1$ in dependence of the magnetic field. The temperatures are ranging up to 10 MeV and $\mu_e$ up to 100 MeV. The required magnetic fields are found to be quite large, between $10^{16}$ G and $10^{18}$ G, but lie below the critical value of $\sim 1.3\cdot 10^{18}$ Gauss for the stability of a magnetised neutron star \citep{Shap_Lai}. From Fig. \ref{pola} we see that for temperatures larger than 4 MeV the magnetic fields should be in the range of $10^{17}$ Gauss and higher. Such large magnetic fields are not seen in observations but predicted to be present during the merger of a binary neutron star system \citep{Price06}. Furthermore, the large magnetic field is just required to be present in the quark phase and can decrease towards the surface of the star. 
\subsection{Polarisation for vanishing temperature}
\label{pol_sec1}
The assumption of zero temperature is usually adopted for describing cold neutron stars or white dwarfs. The number density is given by:
\begin{eqnarray}
n_\mp=\frac{geB}{(2\pi)^2}\sum_{\eta} \int_0^{\sqrt{\mu^2-m^2-2\eta eB}} dp_z.
\end{eqnarray}
The number of Landau levels is limited to $\eta_{max}=\frac{\mu^2-m^2}{2eB}$ where $\mu$ is the electron chemical potential. For $E=\mu$, the electron number densities $n_\mp$ become:
\begin{eqnarray}
n_+&=&\frac{eB}{(2\pi)^2}\sum_{\nu=1}^{\frac{\mu^2-m^2}{2eB}}\sqrt{\mu^2-m^2-2\nu eB}\\
n_-&=&\frac{eB}{(2\pi)^2}\left(\sum_{\nu=1}^{\frac{\mu^2-m^2}{2eB}}\sqrt{\mu^2-m^2-2\nu eB}+\sqrt{\mu^2-m^2}\right).
\end{eqnarray}
The polarisation is consequently:
\begin{eqnarray}
\chi=\frac{n_- - n_+}{n_- + n_+}
&=&\frac{1}{2\left(\sum_{\nu=1}^{\frac{\mu^2-m^2}{2eB}}\sqrt{1-\frac{2\nu eB}{\mu^2-m^2}}\right)+1}.
\label{pol1.1}
\end{eqnarray}
For the case $(\mu^2-m^2)\gg2eB$ the number of occupied Landau levels is large and the sum in eq. (\ref{pol1.1}) can be transferred back to an integration over $\nu$:
\begin{eqnarray}
\int_{1}^{\frac{\mu^2-m^2}{2 e B}} \sqrt{1- \frac{2\nu eB}{\mu^2-m^2}}d\nu&=& \frac{2}{3}\left(1-\frac{2eB}{\mu^2-m^2}\right)^{3/2} \frac{\mu^2-m^2}{2eB}\\
&\sim&\frac{2}{3}\frac{\mu^2-m^2}{2eB}
\end{eqnarray}
which simplyfies the polarisation to:
\begin{eqnarray}
\chi=\frac{1}{2\frac{2}{3}\frac{\mu^2-m^2}{2eB}+1}\sim\frac{3}{2}\frac{eB}{\mu^2-m^2}.
\label{pol_1}
\end{eqnarray}
This relation is easy to understand since an increase of the chemical potential transfers energy to the electrons. Hence, they can overcome the energy gap of $\sqrt{2eB}$ to a new Landau Level with $\eta>0$ where both spin directions are possible which decreases the value for the spin polarisation $\chi$. A look at Fig. \ref{pol1}, where the analytical and numerical results are plotted, shows a very good agreement between these two approaches. The steps which appear in the polarisation curve correspond to the opening of the next Landau level. For sufficiently high chemical potentials the number of Landau levels becomes so high that the effects from the energy quantisation become negligible and the steps disappear.  
\subsection{Polarisation for the non degenerate, massless case ($\mu, m \ll \sqrt{2eB}\ll T$)}
\label{pol_sec2}
The situation which will be discused here can occur in hot proto-neutron stars. Within approximately 15 s after the supernova explosion \citep{Pons99} highly energetic neutrinos propagate from the proto-neutron star interior outwards. Due to their high energy and the large densities the neutrino mean free path is very small. While passing through the neutron star they heat the interior to temperatures of more than 50 MeV \citep{Lattimer04}. Hence, effects from magnetic fields smaller than $10^{17}$ G can be ignored for such high temperatures. For magnetic fields in the range of $10^{16}$ Gauss the condition of $\sqrt{2eB}\ll T$ holds for temperatures larger than 10 MeV. The proto-neutron star temperature drops to $T<$1 MeV very quickly within 50 s \citep{Pons99} and the neutron star becomes transparent to neutrinos. The electron mass and chemical potential are assumed to be negligible in the following. As  $\sqrt{2eB}\ll T$ the number of Landau Levels will be very high and we can again replace the summation over $\eta$ in eq. (\ref{full_dist}) by an integration:
\begin{eqnarray}
n_{\mp}&=&\frac{eB}{(2\pi)^2}\sum_{\eta=0}^{\infty} dp\frac{1}{e^{E/T}+1}
\nonumber\\
&=&\frac{eB}{(2\pi)^2}\int_0^{\infty} dp\int d\eta\frac{1}{e^{E/T}+1}.
\end{eqnarray}
With $\mu, m \ll \sqrt{2eB}$ we have:
\begin{eqnarray}
n_{-}=\frac{eB}{(2\pi)^2}\int dp\left[\int_0^{\infty}d\eta\frac{1}{e^{\sqrt{p^2+2\eta eB}/T}+1}\right],
\end{eqnarray} 
which can be solved with $x=\sqrt{p^2+2\eta eB}/T$:
\begin{eqnarray}
n_{-}&=&\left(\frac{T}{2\pi}\right)^2 \int_0^{\infty} dp\int_{p/T}^{\infty}\frac{1}{e^x+1}dx
\nonumber\\
&=&\frac{3\zeta(3)T^3}{8\pi^2},
\label{negnum}
\end{eqnarray}
where $\zeta(x)$ is the Zeta-function with $\zeta(3)\sim$1.202. The number density $n_{+}$ is:
\begin{eqnarray}
n_+&=&\frac{eB}{(2\pi)^2}\sum_{\eta=1}^{\infty}\int dp\frac{1}{e^{\sqrt{p^2+2\eta eB}/T}+1}\\
&=&\frac{3\zeta(3)T^3}{8\pi^2}-\frac{eB}{4\pi^2}T\ln(2).
\label{posnum}
\end{eqnarray}
and the polarisation becomes:
\begin{eqnarray}
\chi=\frac{n_- - n_+}{n_- + n_+}
=\frac{1}{3\frac{T^2}{eB}\frac{\zeta(3)}{\ln(2)}-1}
\sim\frac{\ln(2)eB}{3T^2\zeta(3)}\mbox{ as $T^2\gg2eB$}.
\label{pol_2}
\end{eqnarray}
From eq. (\ref{pol_2}) we see that for an electron gas with $\mu,m \ll \sqrt{2eB}\ll T$ the polarisation $\chi$ is proportional to $eB/T^2$. The higher the temperature, the more Landau Levels are accessible for the electrons and the electron spin polarisation decreases. A large magnetic field forces the electrons into lower Landau levels. Figure \ref{pol2} shows again a good agreement between the analytical and numerical approach. The values for $\chi$ are very small compared to the previous case (Fig. \ref{pol1}). The number of Landau levels is large and the polarisation goes to zero for high temperatures. The large number of Landau levels is also responsible for the absence of steps. 
\subsection{Polarisation for the non-degenerate, massive case ($\mu=0, m\gg T \gg\sqrt{2eB}$)}
\label{pol_sec3}
In this case we will again discuss high temperatures and weak magnetic fields. The electron mass $m$ is now assumed to be larger than  the temperature and the magnetic field. A star with such conditions could be an old quark star with a colour superconducting quark phase where the electron chemical potential is vanishingly small (e.g. quark matter in the CFL phase). For $m\neq$ 0 the electron number densities have the following form:
\begin{eqnarray}
n_{\mp}=\frac{eB}{(2\pi)^2}\sum_{n}\int dp\frac{1}{e^{\sqrt{p^2+m^2+2\eta eB}/T}+1}.
\end{eqnarray}
Applying the Taylor-expansion for small $x=2eB/T^2$ gives:
\begin{eqnarray}
\nonumber\\
\sqrt{\frac{A^2}{T^2}+\eta x}\simeq\frac{A}{T}+\frac{\eta eB}{AT}
\label{taylor}
\end{eqnarray}
where
\begin{eqnarray}
A=\sqrt{p^2+m^2}.
\end{eqnarray}
The electron number density $n_+$ can be derived by replacing the summation for weak fields with an integration over $\eta$:
\begin{eqnarray}
n_+&\simeq&\frac{eB}{(2\pi)^2} \int dp\left[\int_0^\infty d\eta \frac{1}{e^{\frac{A}{T}+\frac{\eta eB}{AT}}+1}-\frac{1}{e^{A/T}+1}\right]\\
&=&\frac{eB}{(2\pi)^2} \int_0^\infty dp \left[\frac{AT}{eB}\log\left[1+e^{-A/T}\right] \frac{1}{e^{A/T}+1}\right]
\label{numberdensity2}\\
&=&\frac{eB}{(2\pi)^2} \int_m^\infty dA\frac{A^2 T}{eB\sqrt{A^2-m^2}}\log\left[1+e^{-A/T}\right]\\
&-& \frac{eB}{(2\pi)^2} \int_m^\infty dA\frac{A}{\sqrt{A^2-m^2}}\frac{1}{e^{A/T}+1}.
\end{eqnarray}
with
\begin{eqnarray}
\int_0^\infty d\eta \frac{1}{e^{\frac{A}{T}+\frac{\eta eB}{AT}}+1}=\frac{AT}{eB}\log{\left[1+ e^{-A/T}\right]}.
\end{eqnarray}
For small $x$ the Taylor expansion gives $\log(x+1)\sim x$ and therefore for $m \gg T$ one finds:
\begin{eqnarray}
n_+&=&\frac{eB}{(2\pi)^2}\int_m^\infty dA\left(\frac{A^2 T}{eB\sqrt{A^2-m^2}}- \frac{A}{\sqrt{A^2-m^2}}\right)e^{-A/T}\\
& = &\frac{T}{(2\pi)^2}\left(m^2K_0\left[\frac{m}{T}\right] + mTK_1\left[\frac{m}{T}\right] \right) - \frac{eBm}{(2\pi)^2}K_1\left[\frac{m}{T}\right],
\end{eqnarray}
where $K_1$ and $K_0$ are the modified Bessel functions of the second kind. The Bessel functions $K_n$(x) have the following approximate form \citep{Stegun}:
\begin{eqnarray}
K_n(x) \sim \frac{\pi}{\sqrt{2 \pi x}} e^{-x}
\label{approx}
\end{eqnarray}
for $x\gg n$.
One should notice here that the index $n$ in $K_n$(x) does not appear in the approximation anymore. With the number density
\begin{eqnarray}
n_- =  \frac{T}{(2\pi)^2}\left(m^2K_0\left[\frac{m}{T}\right] + mTK_1\left[\frac{m}{T}\right] \right) 
\end{eqnarray}
the polarisation $\chi$ results in:
\begin{eqnarray}
\chi = \frac{1}{\frac{2T}{eB}\frac{mK_0+TK_1}{K_1}-1}
\end{eqnarray}
which becomes for $m \gg T$ using equation (\ref{approx}):
\begin{eqnarray}
\chi = \frac{1}{2(m + T)} \frac{eB}{T} \sim \frac{eB}{2mT}.
\label{pol_3}
\end{eqnarray}
Analytical and numerical results shown in Fig. \ref{pol3} exhibit a small difference from each other. The reason for this deviation is the growing temperature so that at a certain point our assumption of $m \gg T$ is not fulfilled any more. Consequently, the exact numerical result and the approximated analytical solution start to differ from each other. 
\subsection{Polarisation for large magnetic fields ($\mu, m,T \ll\sqrt{2eB}$)}
\label{pol_sec4}
\begin{figure*}
\centering
\subfigure[Polarisation of electrons for $T=0$.]{
 \label{pol1}
\includegraphics[width=0.49\textwidth]{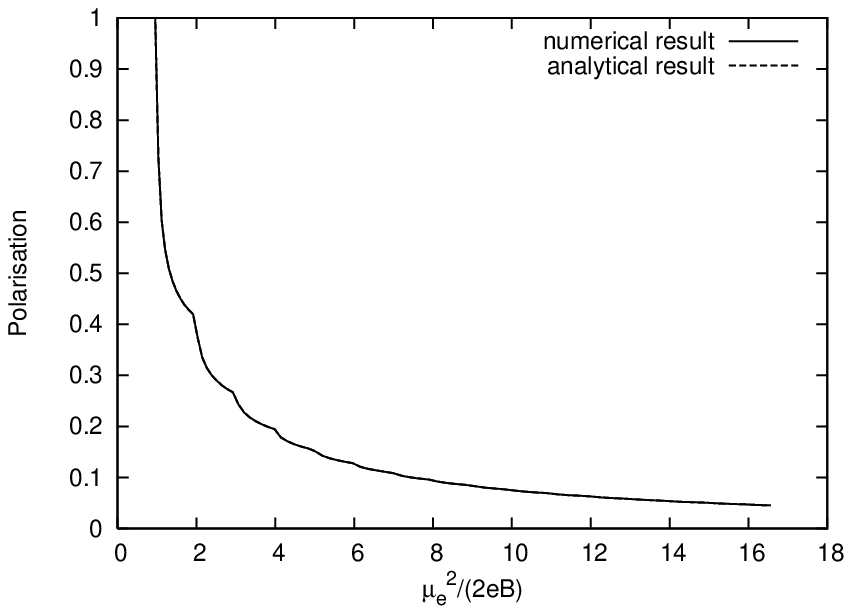} }%
\subfigure[Polarisation for $T^2\gg 2eB \gg m, \mu$.]{
\label{pol2}
\includegraphics[width=0.49\textwidth]{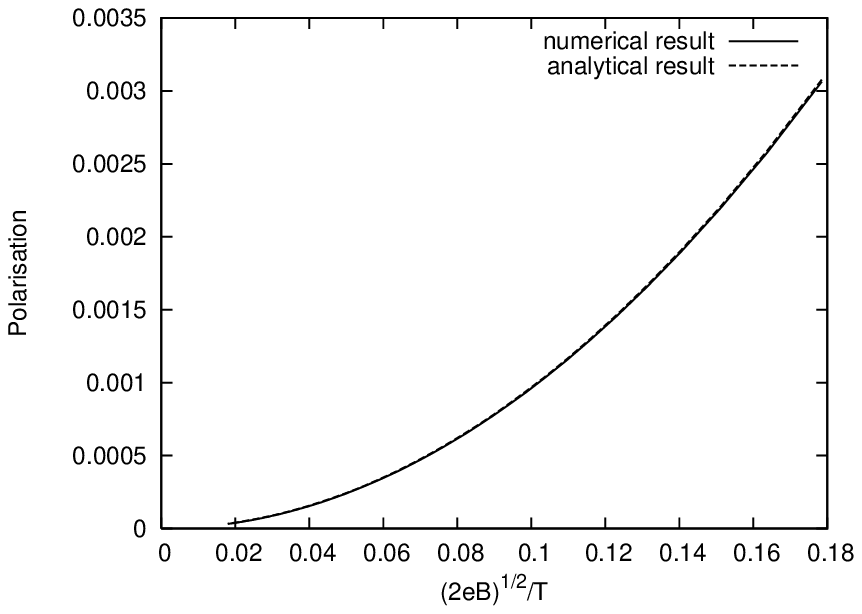}}
\subfigure[Polarisation for $m \gg T \gg\sqrt{2eB}$ and $\mu=0$.]{
\label{pol3}
\includegraphics[width=0.49\textwidth]{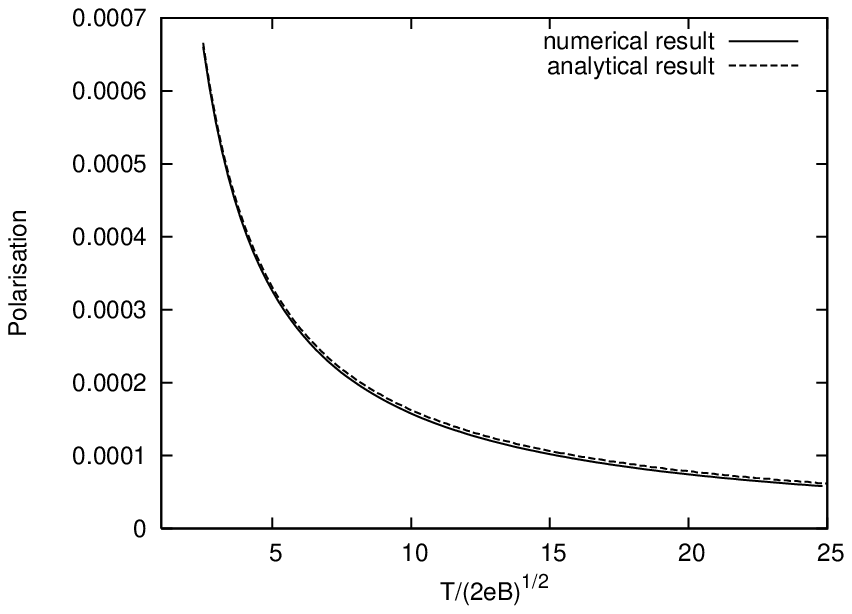}}%
\subfigure[Polarisation for $\mu, m\ll T\ll\sqrt{2eB}$. The magnetic field strength $B$ is $2 \cdot 10^{14}$ Gauss and the temperature is T=0.1 MeV - 2 MeV. The difference between the numerical and the analytical results is max. 10\%.]{
\label{pol4}
\includegraphics[width=0.49\textwidth]{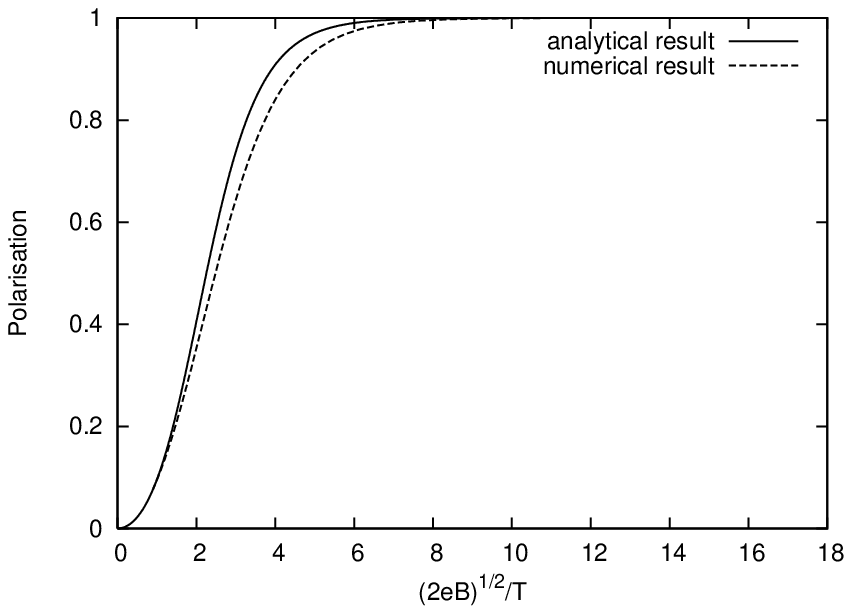}}
\caption{Comparison of the numerical and analytical results for the electron spin polarisation as discussed in Sects. [\ref{pol_sec1}]-[\ref{pol_sec4}]}
\label{pol5}
\end{figure*}

In this section the magnetic field strength is chosen to be much larger than the temperature, the chemical potential as well as the electron mass. This situation is in principle the case discussed in Sect. \ref{pol_sec1} for large magnetic fields and a non-vanishing temperature. The number density integral which we have to solve here is:
\begin{eqnarray}
n_\mp=\frac{eB}{(2\pi)^2}\sum_{\eta}\int_0^{\infty}dp\frac{1}{e^{\sqrt{p^2+2\eta eB}/T}+1},
\end{eqnarray}
where:
\begin{eqnarray}
n_{-}&=&\frac{eB}{(2\pi)^2}\sum_{n=0}^{\infty}\int_0^\infty dp\frac{1}{e^{\sqrt{p^2+2\eta eB}/T}+1}\\
&=&\frac{eB}{(2\pi)^2}\left(T\ln(2)+\sum_{\eta=1}^{\infty}\int_{\sqrt{2\eta eB}/T}^{\infty}\frac{xT dx}{(e^x+1)\sqrt{x^2-\frac{2\eta eB}{T^2}}}\right)
\nonumber\\
&\sim&\frac{eB}{(2\pi)^2}\left(T\ln(2)+\sum_{\eta=1}^{\infty}\int_{\sqrt{2\eta eB}/T}^{\infty}e^{-x} \frac{xT}{\sqrt{x^2-\frac{2\eta eB}{T^2}}}dx \right),
\end{eqnarray}
with $x= \sqrt{p^2+2\eta eB}/T$ and $\sqrt{2\eta eB}/T\gg 1$. The integration over $x$ gives:
\begin{eqnarray}
n_{-}&=&\frac{eB}{(2\pi)^2}\left(T\ln(2)+\sum_{\eta=1}^{\infty}\sqrt{2neB}K_1\left[\frac{\sqrt{2neB}}{T}\right]\right) \label{nminus_1}\\
&=&\frac{eB}{(2\pi)^2}\left(T\ln(2)+\int_{1}^{\infty}d\eta\sqrt{2\eta eB}K_1\left[\frac{\sqrt{2\eta eB}}{T}\right]\right)
\nonumber\\
&=&\frac{eBT\ln(2)}{(2\pi)^2}+\frac{2eB}{(2\pi)^2}K_2\left[\frac{\sqrt{2eB}}{T}\right]T \label{nminus_2}
\end{eqnarray}
where we replaced the summation over $\eta$ again with an integral over d$\eta$. With
\begin{eqnarray}
n_{+}=\frac{2eB}{(2\pi)^2}K_2\left[\frac{\sqrt{2eB}}{T}\right]T
\end{eqnarray}
the polarisation becomes:
\begin{eqnarray}
\chi&=&\frac{1}{1+\frac{4}{ln(2)}K_2\left[\frac{\sqrt{2eB}}{T}\right]}\sim \frac{1}{1+\frac{4}{\ln(2)}\sqrt{\frac{\pi T}{2\sqrt{2eB}}}e^{-\sqrt{2eB}/T}}\\
&\sim& 1-\frac{4}{\ln(2)}\sqrt{\frac{\pi T}{2\sqrt{2eB}}}e^{-\sqrt{2eB}/T}.
\label{pol_4}
\end{eqnarray}
Unfortunately it is quite difficult to approximate the polarisation in eq. (\ref{pol_4}) to a closer expression like  the ones in eqs. (\ref{pol_1}), (\ref{pol_2}) or (\ref{pol_3}). In Fig. \ref{pol4} one can see a sizable deviation of the analytical form from the numerical result starting at $\sqrt{2eB}\sim 2T$ and until $\sqrt{2eB}\sim 7T$. The reason might be the transition from eq. (\ref{nminus_1}) to (\ref{nminus_2}), where we changed the summation to an integration over the Landau levels. The integrand in eq. (\ref{nminus_2}) decreases very fast with $\sqrt{2e\eta B}/T$, consequently only the first Landau levels are significant. The sum over the Landau levels is always larger than the integral over $\eta$. As both, the result of the integration and the summation, are standing in the denominator of the polarisation (see eq. (\ref{pol_4})) the value for $\chi$ is larger for the integration, as seen in Fig. \ref{pol4}. A solution to this problem might be a mathematical more correct way for the transition from the sum to the integral, i.e. the Euler-MacLaurin formalism as used by \citet{Suh}. In table \ref{table_pol} we give a summary of all the analytic solutions for the polarisation discussed here. 
~
\begin{table*}
\caption{Summary of analytic solutions for the polarisation as discussed in Sects. [\ref{pol_sec1}]-[\ref{pol_sec4}].}
\label{table_pol}
\centering
\begin{tabular}{|l|l|l|}
\hline
Case & Polarisation $\chi$ & Approximation\\
\hline
\hline
$T=0$&$\left(2\sum_{\nu=1}^{\frac{\mu_e^2-m^2}{2eB}}\sqrt{1-\frac{2\nu eB}{\mu_e^2-m^2}}+1\right)^{-1}$&${3eB}/{2\mu_e^2}$ for $\mu_e\gg\sqrt{2eB},m$\\
$T \gg 2eB \gg \mu, m$&$\left(3\frac{T^2}{eB}\frac{\zeta(3)}{\ln{2}}-1\right)^{-1}$&${eB\ln{2}}/{3\zeta(3)T^2}$ as $T^2 \gg 2eB$\\
$m \gg T \gg 2eB, \mu_e=0$&$\left(\frac{2T}{eB}\frac{mK_0[m/T]+T K_1[m/T]}{K_1[m/T]}-1\right)^{-1}$&${eB}/{2mT} $ for $m \gg T$\\
$2eB \gg T \gg m, \mu$&$\left(1+\frac{4}{ln(2)}\sqrt{\frac{\pi T}{4eB}}e^{-\sqrt{2eB/T}} \right)^{-1}$&--\\
\hline
\end{tabular}
\end{table*}
\section{Calculation of Pulsar velocities}
\label{calc_pv}
\subsection{Heat Capacity}
\label{h_cap}
The specific heat capacity per volume c$_v$ can be calculated from the internal energy density $\epsilon$ by:
\begin{eqnarray}
c_V=\left(\frac{\partial\epsilon}{\partial T}\right)_{V,N}.
\label{hc_eps}
\end{eqnarray}
The energy density of non-interacting relativistic fermions is given by:
\begin{eqnarray} 
\epsilon=\frac{gT^2\mu^2}{4 }+\frac{7}{120}{g\pi^2}(T)^4+\frac{g\mu^4}{8\pi^2},
\label{en_den_ferm}
\end{eqnarray}
with $g$ being the degeneracy factor. Quark interactions can be taken into account by perturbative QCD calculations to first order in the strong coupling constant $\alpha_s= g^2/(4\pi)$ (see e.g. \citet{Glendenning}):
\begin{eqnarray}
\epsilon_q=\frac{7g}{120}\pi^2T^4\left(1-\frac{50\alpha_s}{21\pi}\right)+g\left(\frac{T^2 {\mu_q}^2}{4}+\frac{{\mu_q}^4}{8\pi^2}\right)\left(1-\frac{2\alpha_s}{\pi}\right).
\label{epsilon_2}
\end{eqnarray}
Consequently the heat capacity of quarks becomes:
\begin{eqnarray}
c_q=\left(\frac{\partial\epsilon}{\partial T}\right)_V=9\mu^2T\left(1-\frac{2 \alpha_s}{\pi}\right)+\frac{21}{5}\pi^2T^3\left(1-\frac{50\alpha_s}{21\pi}\right)
\label{cap_q}
\end{eqnarray}
with the degeneracy factor $g=2\times 3 \times 3=18$. Proto neutron stars can have temperatures up to 50MeV with quark chemical potentials in the range of (400 - 500) MeV. The chemical potentials for the electrons are much lower. With $\mu_q\gg\mu_e$ the heat capacities for quarks $c_q$ are much higher than the ones for the electrons $c_e$. 
The total heat capacity is then to a good approximation:
\begin{eqnarray}
c_{total} = c_{q}=9{\mu_q}^2T\left(1-\frac{2\alpha_s}{\pi}\right).
\label{cap_q2}
\end{eqnarray}
\subsection{Pulsar acceleration}
\label{acceleration}
\begin{figure}
{\centering \includegraphics[width=0.34\textwidth,angle=270]
{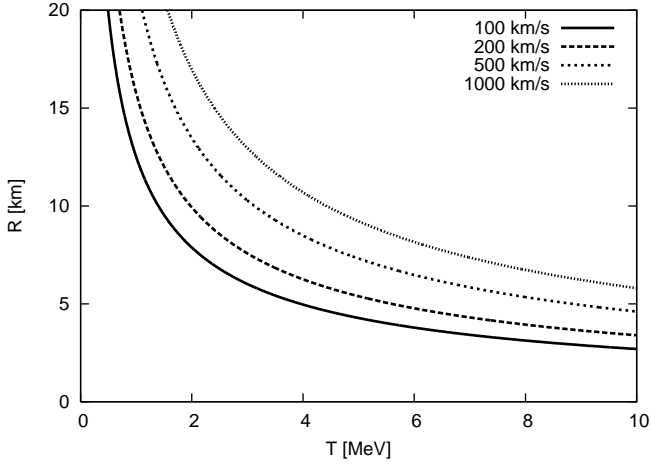} \par}
\caption{Radius and temperature dependence of the kick velocity for $\mu_q=400$ MeV, $\alpha_s=0.5$, M$_{ns}$ = 1.4 M$_\odot$ and $\chi=1$. }
\label{just_vel}
\end{figure}

As already discussed, the kick mechanism under investigation is based on anisotropically emitted neutrinos which accelerate the neutron star in the opposite direction of their emission. The amount of acceleration depends on the polarisation of the electron spin and the neutrino momenta, respectively. Assuming a negligible neutrino mass we can write for the kick velocity:
\begin{eqnarray}
dv=\frac{\chi}{M_{ns}}\frac{4}{3}\pi R^3 \epsilon_{q\beta}dt.
\label{dvns}
\end{eqnarray}
whereas $\epsilon_{q\beta}$ is the neutrino emissivity, $R$ the radius of the neutrino emitting quark phase and $\chi$ the fraction of polarised neutrinos. Using $\epsilon_{q\beta} = - d\epsilon/dt$ the velocity of the neutron star will depend on the energy density in the following way:
\begin{eqnarray}
v&=&\frac{4}{3}\pi R^3\frac{\chi}{M_{ns}}\left( \epsilon(t_0)-\epsilon(t_f)\right)= \frac{4}{3}\pi R^3\frac{\chi}{M_{ns}}\Delta \epsilon
\label{vel_gen}\\
&\sim & 700 \frac{km}{s}\chi\left(\frac{\Delta \epsilon}{{\rm MeV fm}^{-3}}\right)\left(\frac{R}{10{\rm km}}\right)^3\left(\frac{1.4 M_\odot}{M_{ns}}\right).
\label{acc_mev}
\end{eqnarray}
In principle the difference in energy densities in eq. (\ref{vel_gen}) could be produced in various ways in quark matter due to the appearance of colour-superconducting phases \citep{Lavagno07,Noronha07}. In the following we will concentrate on $\Delta\epsilon$ originating from temperature decay. The right-hand side in eq. (\ref{dvns}) can be replaced by an integration over temperature using eq. (\ref{hc_eps}):
\begin{eqnarray}
dv = \frac{4}{3}\pi R^3\epsilon_{q\beta}\frac{\chi}{M_{ns}}\left(-\frac{c_q}{\epsilon_{q \beta}}\right)dT
= -\frac{4}{3}\pi R^3\frac{\chi}{M_{ns}}c_q dT.
\label{velocity1}
\end{eqnarray}
Hence,
\begin{eqnarray}
v &=& \frac{2}{3}\pi R^3\frac{\chi}{M_ns}9\left(1-\frac{2\alpha_s}{\pi}\right){\mu_q}^2(T_0^2-{T_f}^2)\\
&\sim& 40 \frac{{\rm km}}{{\rm s}}\left(\frac{\mu_q}{400{\rm MeV}}\right)^2\left(\frac{T_0}{\rm MeV}\right)^2\left(\frac{R}{10{\rm km}}\right)^3\frac{1.4 M_\odot}{M_{ns}},
\label{velocity2}
\end{eqnarray}
where $T_0$ and $T_f$ are the initial and the final temperatures, respectively. In these calculations we neglect any effects from the magnetic field on the energy density of quarks and electrons. Considering magnetic field strengths in the range of $10^{16}-10^{17}$ Gauss and temperatures around 1-10 MeV the phase space of the electrons will differ from the one of relativistic fermions. However, the electron contribution to the total energy density is small and quarks can be considered as a free gas of fermions due to their large chemical potential. The situation might change when considering gapped quark matter. From eq.  (\ref{velocity1}) we can see that the neutrino emissivity completely drops out of the expression for the kick velocity. The crucial input for the acceleration is the thermal energy stored in the medium. If the emissivity is small the neutron star cools slower for a longer time period. If the emissivity is large, the cooling is quick. The velocity estimate in eq. (\ref{velocity2}) shows that temperatures smaller than 1 MeV will not give observable velocities. A proto-neutron star cools down to temperatures below 1 MeV during the first minutes \citep{Lattimer04}. Therefore the final temperature $T_f$ can be neglected. For  $\alpha_s=0.5, \mu_q=400$ MeV, $M_{ns}=1.4 M_\odot$ and $\chi=1$, the dependency of the kick velocity on the temperature and the radius of the neutrino emitting quark phase is plotted in Fig. \ref{just_vel}.
The higher the temperature of the quark phase the higher is the neutrino energy, $E_\nu=3.15T$, which results in smaller radii $R$ for a given velocity. For an initial temperature of $T>5$ MeV a kick of 1000 km s$^{-1}$ is possible for $R <$10 km. From Fig. \ref{pola} we find that these temperatures require magnetic fields in the range of $B \sim 10^{17}$ G to achieve full polarisation for different electron chemical potentials. These magnetic fields seem to be very high compared to the observed ones of approximately $B \sim 10^{12}$ G. Nevertheless, it is possible that these high magnetic fields are present in the interior of the neutron star and might be substantially smaller at the surface.
 \section{Neutrino mean free paths}
\label{neutrino_mean}
For hybrid stars or strange stars with a nuclear crust we have to consider at least four neutrino interaction processes with the medium: The absorption of neutrinos in quark matter (d + $\nu_e\longrightarrow$u + e$^-$) and neutron matter (n + n + $\nu_e\longrightarrow$n + p + e$^-$) as well as the the scattering processes (q + $\nu\longrightarrow$q + $\nu$ and n + $\nu\longrightarrow$n + $\nu$). We will calculate these mean free paths following closely \citet{Iwamoto81}. Yet, we have to keep in mind that Iwamoto did not consider magnetic field effects and assumed degenerate electrons which will not be the case for high temperatures. For high temperatures one can apply the results of \citet{Burrows80} who arrives at a temperature dependence of the Urca emissivity of $T^7$ instead of the $T^6$ behaviour for non-degenerate electrons. 
\subsection{Absorption of non degenerate neutrinos in quark matter}
For the case of absorption of non-degenerate neutrinos in quark matter 
\begin{eqnarray}
d + \nu_e \longrightarrow u + e^{-}.
\label{absorption}
\end{eqnarray}
the neutrino mean free path $l^q_{abs}$ can be calculated for a given temperature and neutrino energy by the following expression \citep{Iwamoto81}:
\begin{eqnarray}
\frac{1}{l_{abs}^q}=\frac{4}{\pi^4}\alpha_s G_F^2\cos^2\Theta_c p_F(d) p_F(u) p_F(e^{-})\left(\frac{E_\nu^2+(\pi T)^2}{1+e^{-E_\nu/T}}\right),
\label{1overlabs}
\end{eqnarray}
where $G_F$ is the weak coupling constant and the Cabbibo angle is $\cos^2\theta$=0.948. The Fermi momenta for the electron, the up and down quarks $p_F$(i) with $i=e^{-},u,d$ can be expressed by their respective chemical potential using: 
\begin{eqnarray}
\mu_{u,d} =\left( 1+ \frac{2 \alpha_s}{3\pi}\right) p_{F}(u,d) = \mu_q,
\label{qfm}
\end{eqnarray}
and
\begin{eqnarray}
\mu_{e}= p_{F}(e) \ll \mu_q.
\end{eqnarray}
The strong coupling constant $\alpha_s$ from quark interactions appears necessarily \citep{Iwamoto81} whereas the prefactor $4/\pi^4$ in eq. (\ref{1overlabs}) differs from the one used by \citet{Iwamoto81}, $16/\pi^4$, as he applied a modified definition of $\alpha_s=g^2/16\pi$. With $\alpha_s=g^2/4\pi=0.5$ the neutrino mean free path becomes: 
\begin{eqnarray}
l_{abs}^q &\sim& 570{\rm km} \left(1+e^{-E_\nu/T}\right) \left(\frac{\mu_e}{\rm MeV}\right)^{-1}\left(\frac{\mu_q}{\rm 400MeV}\right)^{-2}
\nonumber\\
&&\times \left(\frac{\pi^2 T^2+E_\nu^2}{\rm MeV^2}\right)^{-1}\\
&\sim& 30{\rm km}\left(\frac{\mu_e}{\rm MeV}\right)^{-1}\left(\frac{\mu_q}{\rm 400MeV}\right)^{-2}\left(\frac{T}{\rm MeV}\right)^{-2},
\label{result1}
\end{eqnarray}
setting $E_\nu= 3.15 T$. One should notice that the relation between $l_{abs}^q$ and $\mu_e$ in equation (\ref{result1}) is not quite intuitive. The absorption reaction in eq. (\ref{absorption}) should be suppressed for a high number of electrons present that is for a high electron chemical potential. Hence, one would guess that $l_{abs}^q$ becomes smaller with decreasing $\mu_e$. In contrast the scaling of $l_{abs}^q$ with temperature is more natural. The larger the temperature the higher is the neutrino energy which increases the interaction rate.
\subsection{Scattering of non degenerate neutrinos in quark matter}
For the scattering of non degenerate neutrinos in quark matter and neglecting quark-quark interactions \citet{Iwamoto81} gives:
\begin{eqnarray}
l_{scat}^i&=&\frac{20}{C_{Vi}^2+C_{Ai}^2}\frac{1}{n_i\sigma_0} \left(\frac{m_e}{E_\nu}\right)^2\left(\frac{p_F(i)}{E_\nu}\right)
\nonumber\\
&=&\frac{20}{C_{Vi}^2+C_{Ai}^2}\frac{\pi^2}{\mu_q^2 \sigma_0 m_e}\left(\frac{m_e}{E_\nu}\right)^3,
\label{li}
\end{eqnarray}
whereas $i$ is the quark component flavour, $n_i$ is its number density and $\sigma_0= 4m_e^2 G_F^2/\pi$. The parameters $C_{Vi}$ and $C_{Ai}$ are the vector and axial coupling constants and can be found in the article by \citet{Iwamoto81}. The total scattering mean free path is now given by:
\begin{eqnarray}
l_{scat}^{tot}&=&\left(\sum_{i=u,d,s}\frac{1}{l_{scat}^{i}}\right)^{-1}
\label{add_mfp}\\
&=&\frac{20}{2C_{Vd}^2+2C_{Ad}^2+C_{Vu}^2+ C_{Au}^2}\frac{\pi^2}{\mu_q^2\sigma_0 m_e}\left(\frac{m_e}{E_\nu}\right)^3\\
&\sim& 1370{\rm km}\left(\frac{\mu_q}{\rm 400MeV}\right)^{-2}\left(\frac{E_\nu}{\rm MeV}\right)^{-3}
\nonumber\\
&\sim& 40 {\rm km}\left(\frac{\mu_q}{\rm 400MeV}\right)^{-2}\left(\frac{T}{\rm MeV}\right)^{-3},
\label{result2}
\end{eqnarray}
using again $E_\nu \simeq 3.15 T$. Since the electrons do not play an important role for elastic neutrino-quark scattering their chemical potential does not occur in the final expression for the mean free path. For fixed quark chemical potentials the interaction rate is determined by the neutrino energy. 
\subsection{Absorption and Scattering of non degenerate neutrinos in hadronic matter}
For the absorption and scattering in degenerate neutron matter we refer again to \citet{Iwamoto81}:
\begin{eqnarray}
l_{scat}^n&=&\left(\frac{3}{32}\left(1+3g_A^2 \right)n_n \sigma_0 \left(\frac{E_\nu}{m_e}\right)^2\left(\frac{T}{E_F(n)}\right)\right)^{-1}\\
&=&\left(\frac{1}{4\pi^3}\left(1+3g_A^2\right)p_{Fn} G_F^2 E_\nu^2 m_n T\right)^{-1}\\
&\sim& 100 {\rm km}\left(\frac{E_\nu}{\rm MeV}\right)^{-2}\left(\frac{T}{\rm MeV}\right)^{-1} \left(\frac{p_{Fn}}{330 {\rm MeV}}\right)^{-1}\\ &\sim& 10{\rm km}\left(\frac{T}{\rm MeV}\right)^{-3}\left(\frac{p_{Fn}}{330 {\rm MeV}}\right)^{-1}
\label{result3}
\end{eqnarray}
where the neutron Fermi energy is defined by $E_F(n)=p_{Fn}^2/2m_n$. For evaluating the mean free path in eq. (\ref{result3}) we set $g_A = 1.25$ and again $E_\nu=3.15 T$. For the neutrino absorption by neutrons we use:
\begin{eqnarray}
l^n_{abs}&=& \frac{45 {\rm km}}{(y^4+10\pi^2y^2+9\pi^4)}\left(\frac{T}{\rm 10 MeV}\right)^{-4}\left(\frac{n_b}{n_0}\right)^{-2/3}
\label{midresult4}\\
&\sim& 230{\rm km}\left(\frac{T}{\rm MeV}\right)^{-4} \left(\frac{p_{Fn}}{\rm 340 MeV}\right)^{-2}
\label{result4}
\end{eqnarray}
with $y=E_\nu/T$ and $n_b=n_n$. For $E_\nu \gg T$ eq. (\ref{midresult4}) gives:
\begin{eqnarray}
l^n_{abs}&\simeq& 45 {\rm km} \left(\frac{E_\nu}{\rm 10 MeV}\right)^{-4}\left(\frac{p_{Fn}}{\rm 340 MeV}\right)^{-2}
\end{eqnarray}
\subsection{Discussion of the kick velocity considering neutrino mean free paths}
\begin{figure}
{\centering \includegraphics[width=0.32\textwidth, angle=270]
{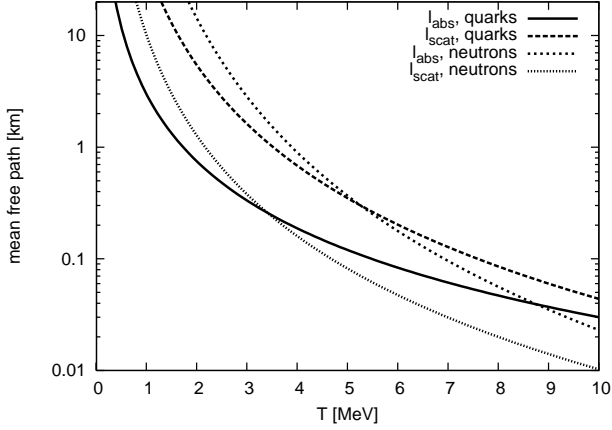} \par}
\caption{Neutrino mean free paths in quark matter and neutron matter for $E_\nu=3.15T$, $\mu_q=400$ MeV, $\alpha_s$=0.5, $n_n=n_0$ and $\mu_e$=10 MeV.}
\label{mfp}
\end{figure}

\begin{figure}
{\centering \includegraphics[width=0.34\textwidth, angle=270]{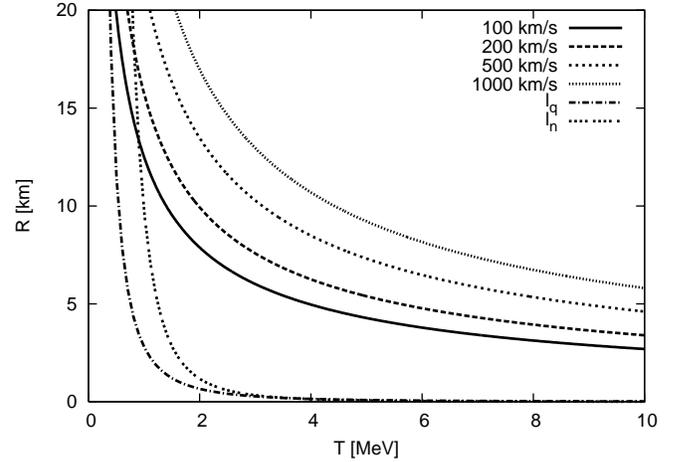} \par}
\caption{Neutrino mean free paths in quark matter and neutron matter for $E_\nu=3T$, $\mu_q=400$ MeV, $\alpha_s=0.5$, $n_n=0.16$  fm$^{-3}$ and $\mu_e=10$ MeV with kick velocities for $M_{ns}=1.4 M_{\odot}$ and full electron spin polarisation.}
\label{vel_mfp}
\end{figure}

\begin{figure*}
\centering
\subfigure[Strong coupling constant $\alpha_s=0.3$.]{
 \label{var_alpha1}
\includegraphics[width=0.34\textwidth, angle=270]{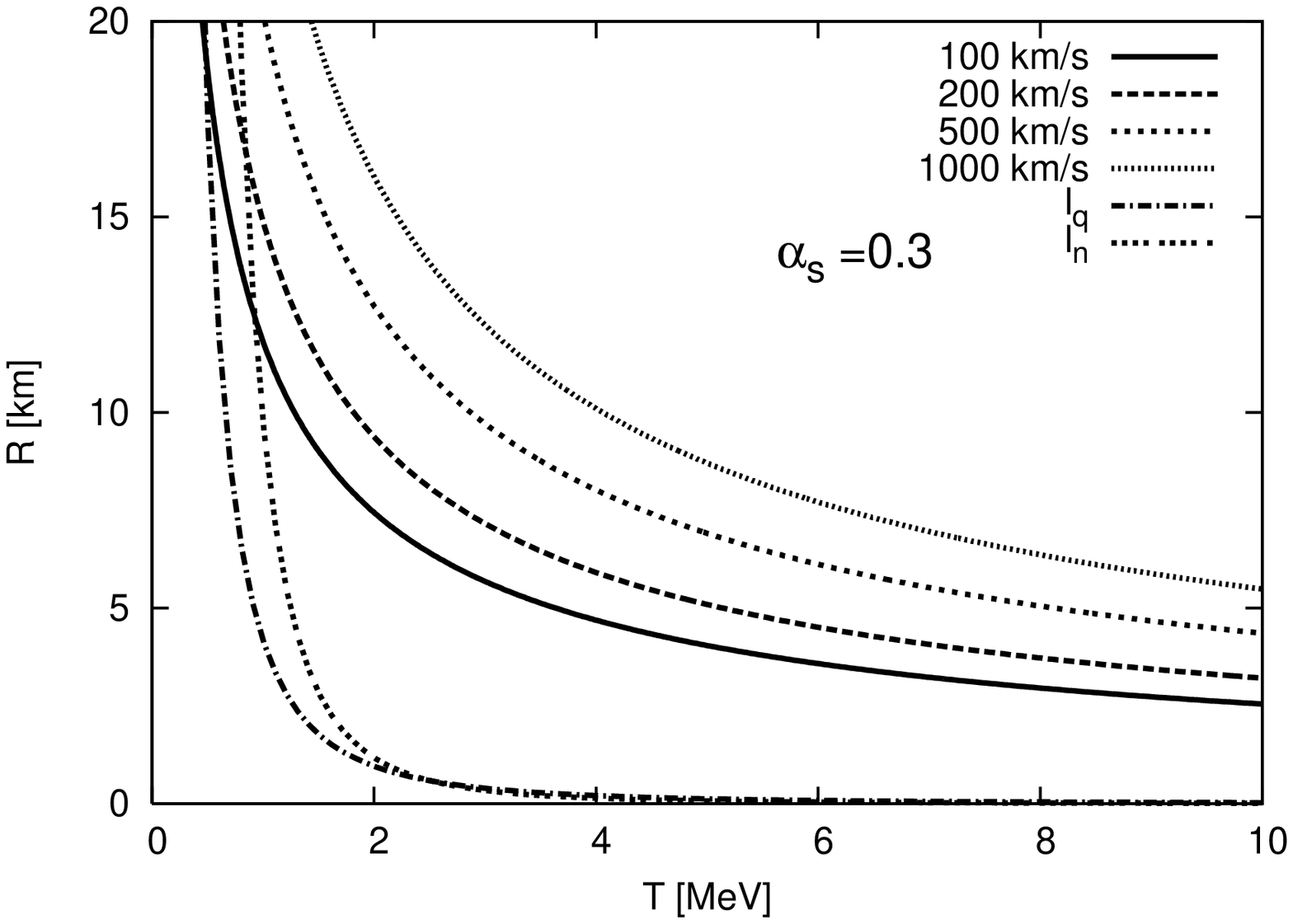}}
\subfigure[Strong coupling constant $\alpha_s=0.7$.]{
\label{var_alpha2}
\includegraphics[width=0.34\textwidth, angle=270]{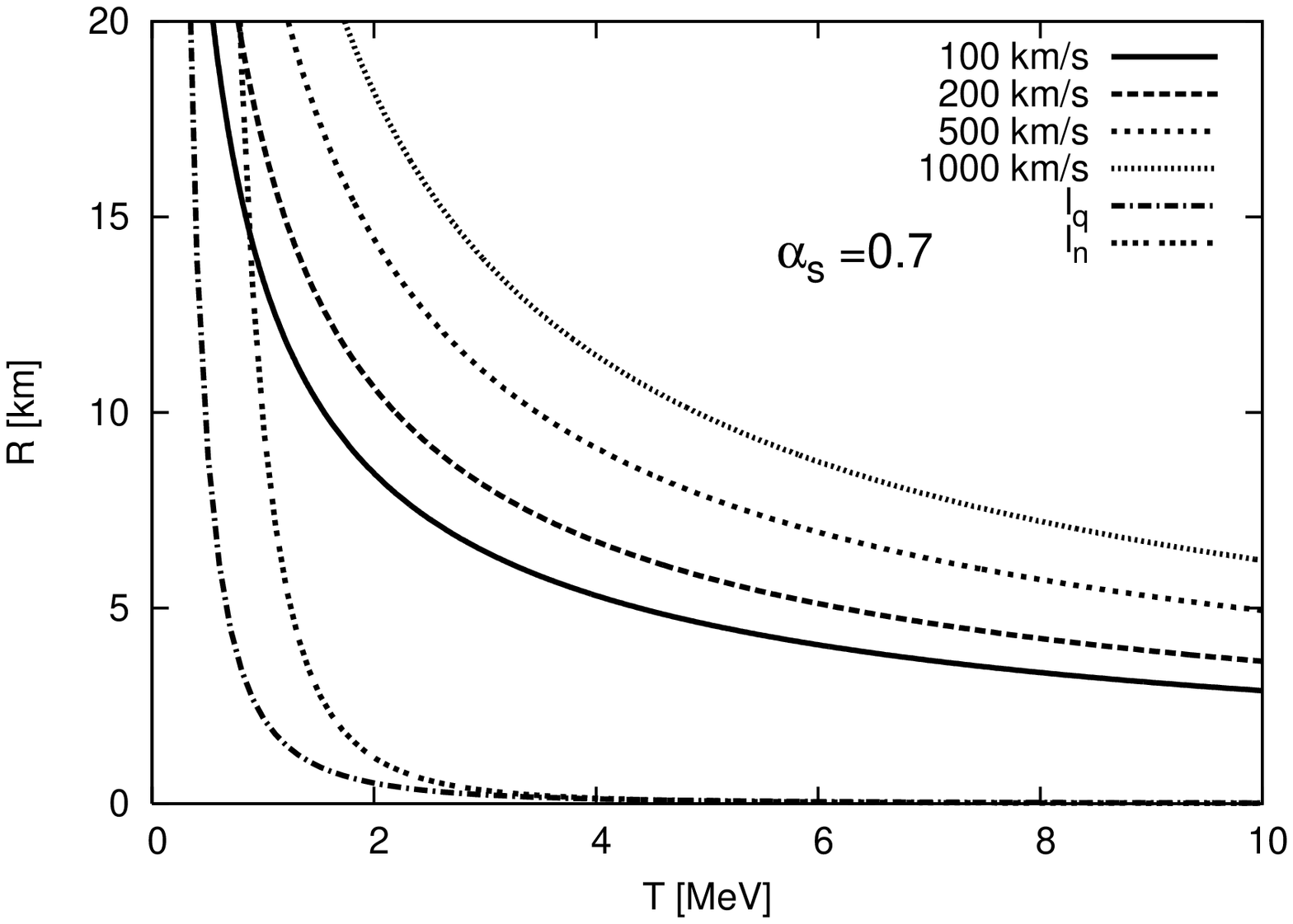}}
\caption{Kick velocities and neutrino mean free paths in quark matter and neutron matter as in Fig. \ref{vel_mfp} for different values of $\alpha_s$.}
\label{var_alpha}
\end{figure*}

\begin{figure*}
\centering
\subfigure[Quark chemical potential $\mu_q=300$MeV]{
 \label{var_muq1}
\includegraphics[width=0.34\textwidth, angle=270]{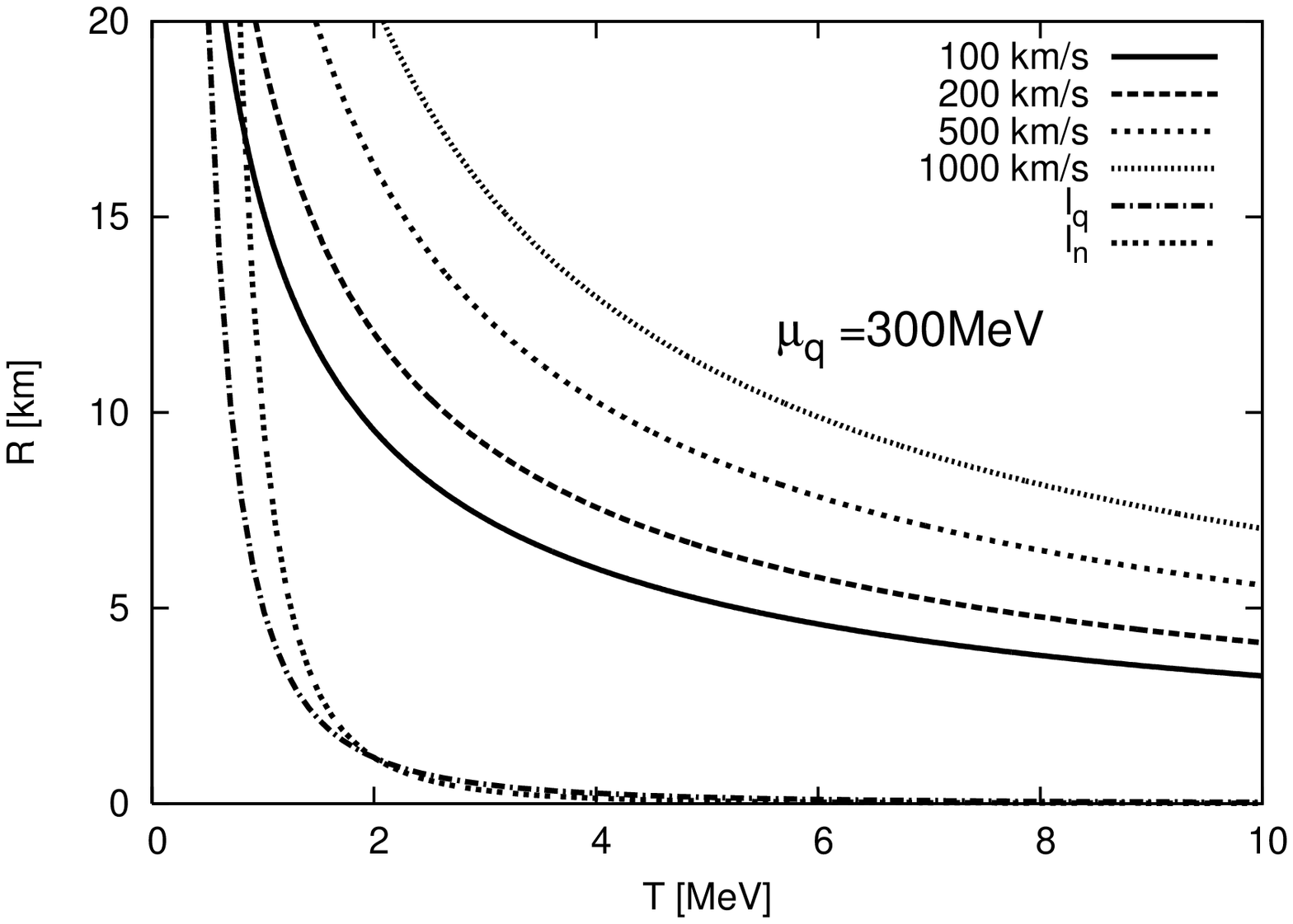}}
\subfigure[Quark chemical potential $\mu_q=500$MeV]{
\label{var_muq2}
\includegraphics[width=0.34\textwidth, angle=270]{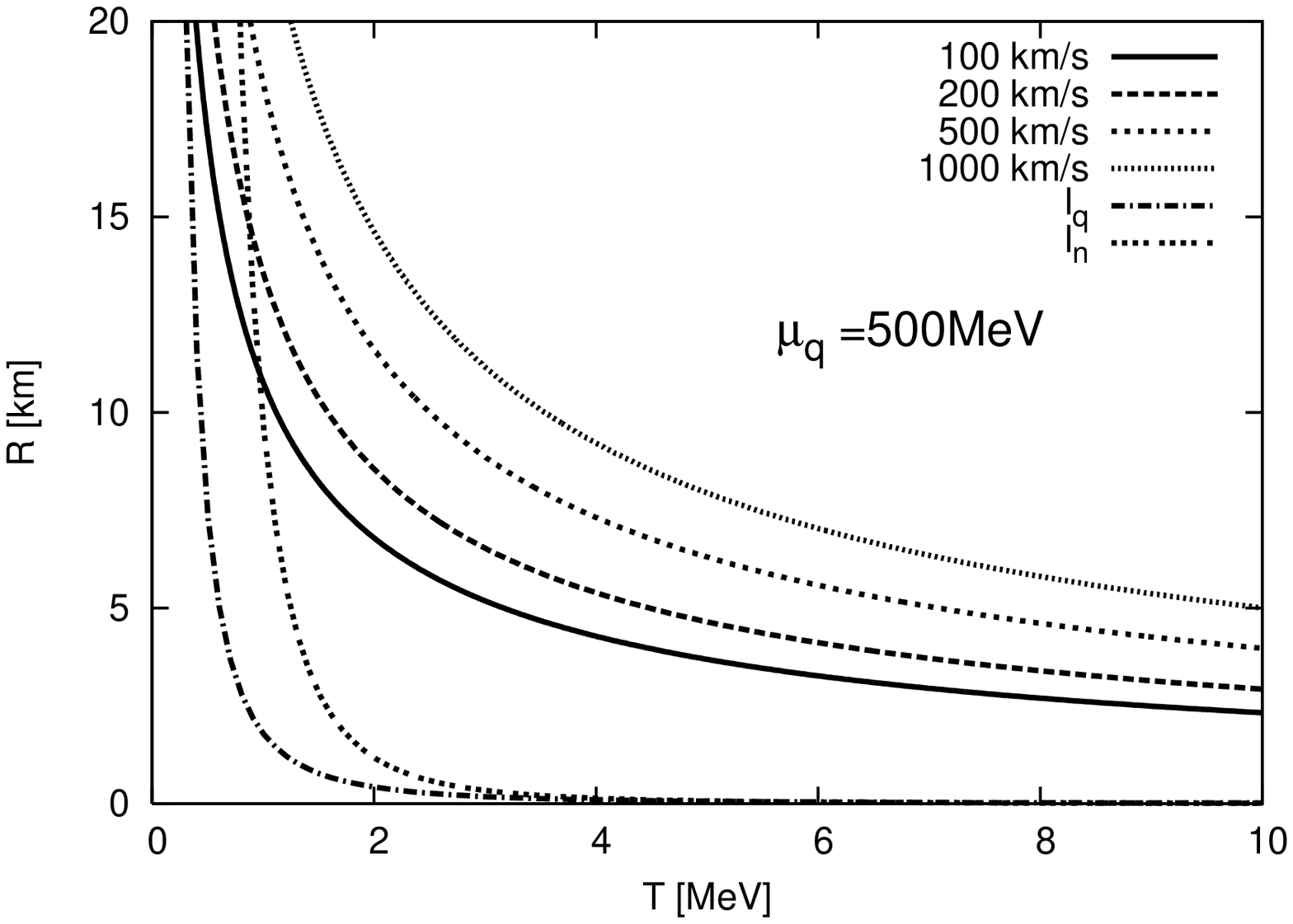}}
\caption{Kick velocities and neutrino mean free paths in quark matter and neutron matter as in Fig. \ref{vel_mfp} for different values of $\mu_q$.}
\label{var_muq}
\end{figure*}

\begin{figure*}
\centering
\subfigure[Quark chemical potential $\mu_q=500$MeV and $\alpha_s=0.3$.]{
 \label{0.3_500_10}
\includegraphics[width=0.34\textwidth, angle=270]{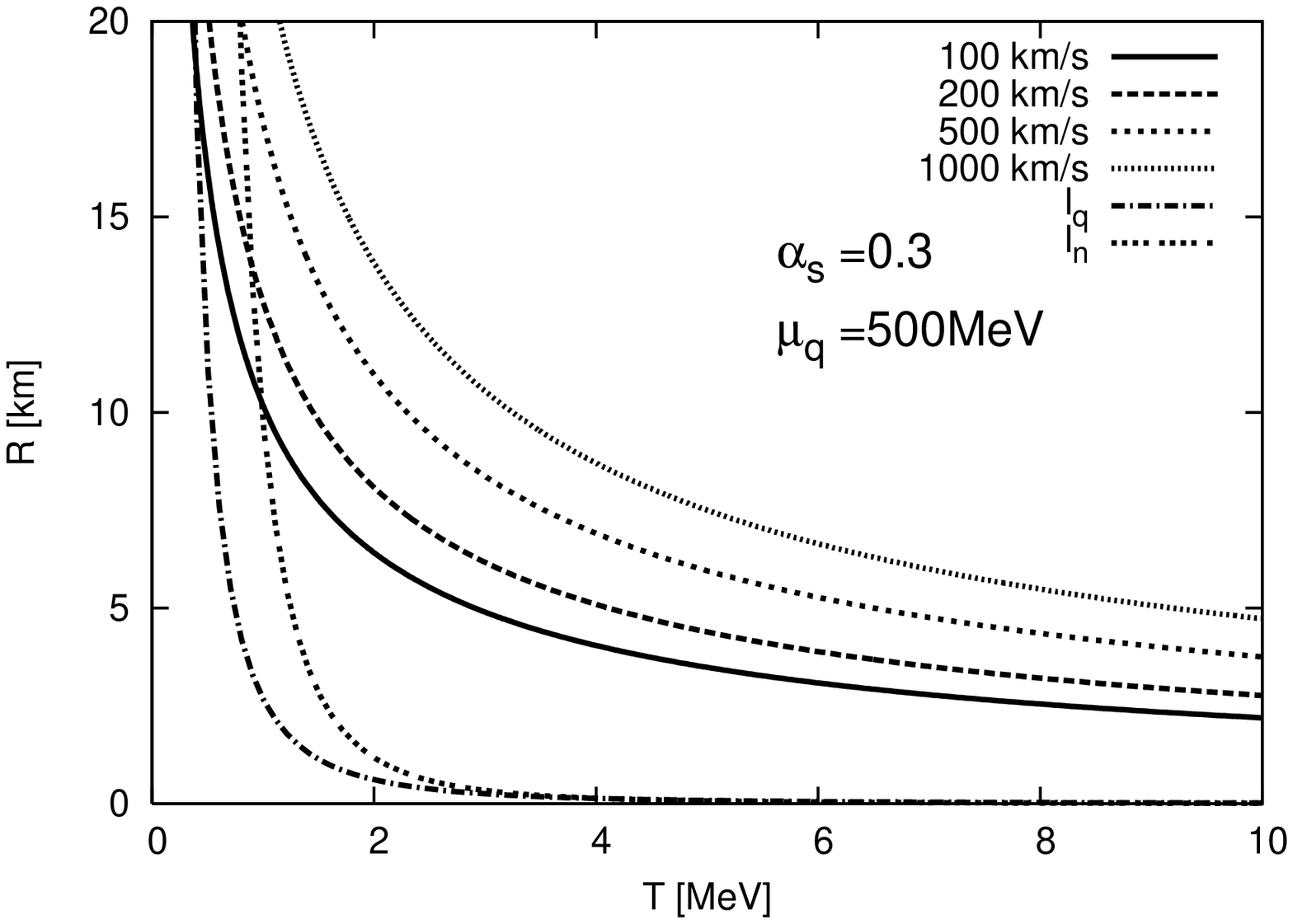}}
\subfigure[$\alpha_s=0.5$ and electron chemical potential $\mu_e\sim T$.]{
\label{0.3_500_T}
\includegraphics[width=0.34\textwidth, angle=270]{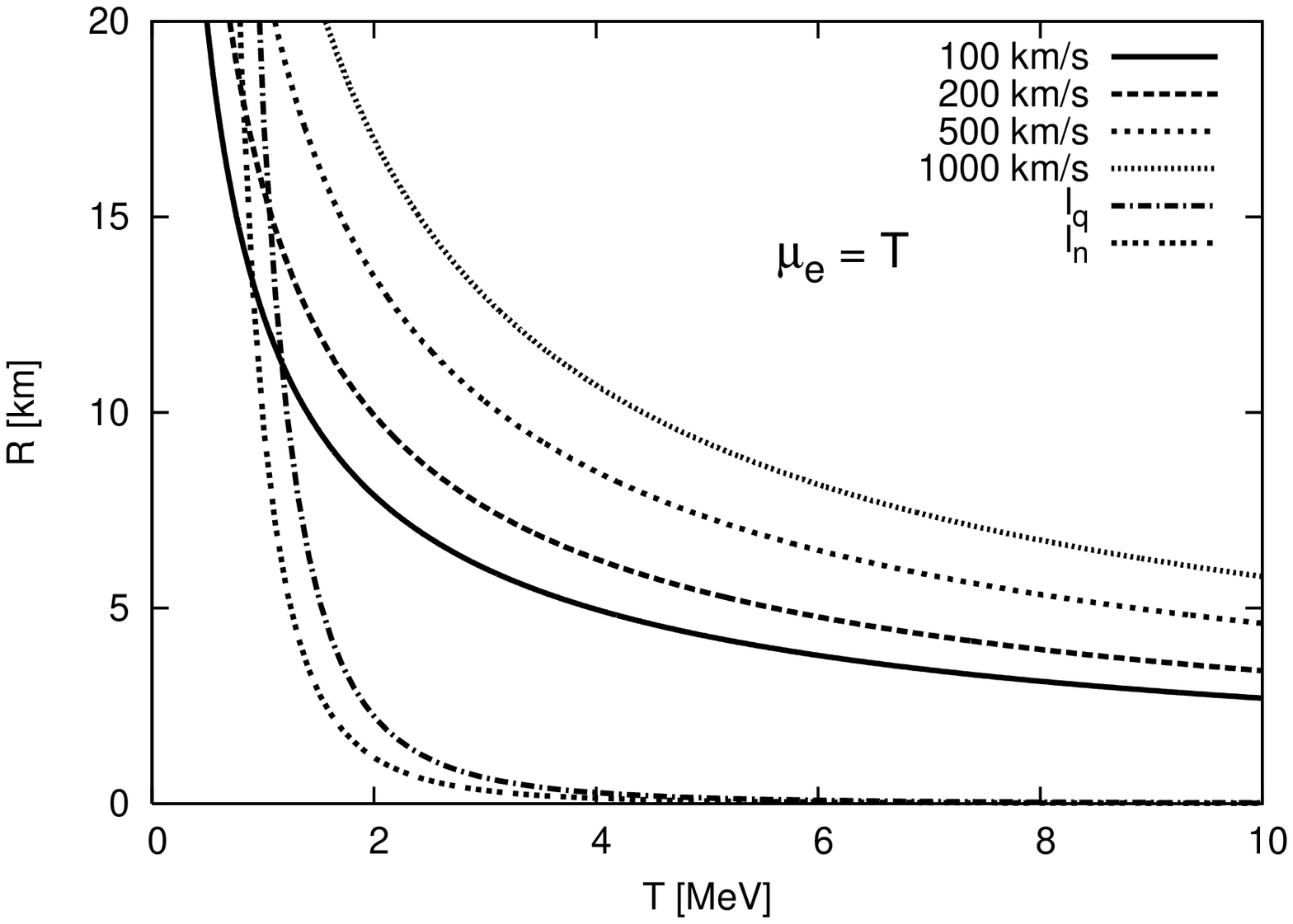}}
\caption{Kick velocities and neutrino mean free paths in quark matter and neutron matter as in Fig. \ref{vel_mfp}. In \ref{0.3_500_10} the strong coupling constant is set to $\alpha_s=0.3$, the electron chemical potential $\mu_e$ is 10 MeV and $\mu_q=500$ MeV. In \ref{0.3_500_T} the electron chemical potential is set to be equal to the temperature $T$, $\mu_q=400$ MeV and $\alpha_s=0.5$.}
\label{muq500}
\end{figure*}

At first sight the mean free paths in the equations (\ref{result1}), (\ref{result2}), (\ref{result3}) and (\ref{result4}) seem to be quite large. In Sect. \ref{acceleration} it was found that the required temperatures for the observed pulsar kicks should be around 5 MeV. For this temperature the mean free paths decrease drastically to the order of $\sim$100 m as can be seen from table \ref{irina_tab} and Fig. \ref{mfp} where we plot the neutrino mean free paths in dependence of the temperature.
\begin{table*}
\caption{Neutrino mean free paths for absorption and scattering processes in quark matter as well as in neutron matter with $n_0 = 0.16$ fm$^{-3}$ and the neutron Fermi momentum $p_{Fn} =\left(3 \pi^2 n_n\right)^{1/3} \simeq 330 \left(n_n/n_0\right)^{1/3}$ MeV .}
\label{irina_tab}
\centering
\begin{tabular}{|l|l|l|l|l|}
\hline
Medium & Process &$E_\nu=3.15T, \alpha_s=0.5$ &T=5MeV, $\mu_e$=10MeV\\
\hline
\hline
Quark matter&Absorption&$l_{abs}\sim$ 30 km $\left(\frac{T}{{\rm MeV}}\right)^{-2}\left(\frac{\mu_q}{\rm  400 MeV}\right)^{-2}\left(\frac{\mu_e}{\rm MeV}\right)^{-1}$&$\sim$ 120 m for $\mu_q$=400 MeV\\ 
Quark matter&Scattering&$l_{scat}\sim$ 40 km $\left(\frac{\mu_q}{\rm 400 MeV}\right)^{-2}\left(\frac{T}{\rm MeV}\right)^{-3}$&$\sim$ 350 m for $\mu_q$=400 MeV\\
Neutron matter&Absorption&$l_{abs}\sim$ 230 km $\left(\frac{T}{\rm MeV}\right)^{-4}\left(\frac{p_{Fn}}{\rm 330 MeV} \right)^{-2}$&$\sim$ 370 m for $n_n=n_0$\\
&&&$\sim$ 130 m for $n_n \simeq 5 n_0$\\
Neutron matter&Scattering&$l_{scat}\sim$10 km $\left(\frac{T}{\rm MeV}\right)^{-3}\left(\frac{p_{Fn}}{\rm 340 MeV} \right)^{-1}$&$\sim$ 80 m for $n_n=n_0$\\
&&&$\sim$ 50 m for $n_n \simeq 5 n_0$\\
\hline
\end{tabular}
\end{table*}
More modern calculations of the neutrino mean free path in colour flavour locked quark matter were done e.g. by \citet{Reddy02}. The authors applied for the neutrino mean free paths in nuclear matter results obtained by \citet{PraLat} and by \citet{Iwamoto81} for unpaired quark matter. The neutrino energy was set to $E_\nu$=$\pi T$ and the baryon density is $n_b = 5n_0$ which corresponds to a quark chemical potential of $\mu\sim 400$ MeV.\\
For neutrino mean free paths in unpaired quark matter \citet{Reddy02} obtain the same results as in table \ref{irina_tab}. Large differences arise comparing the neutrino mean free paths for absorption and scattering in neutron matter. For a temperature of 5 MeV, the authors get for the scattering a mean free path of 200 m, which is higher than our result (see table \ref{irina_tab}), and a mean free path for neutrino absorption of 2 m.  The large difference in the neutrino mean free path for absorption is due to the different processes applied. While \citet{PraLat} considered the direct nucleon Urca process for $n_B \sim 5n_0$ we applied the modified Urca process assuming a neutron crust where the density is not high enough for the direct Urca process to occur. For the next discussions we will combine the neutrino mean free paths for quark matter as well as neutron matter using eq. (\ref{add_mfp}):
\begin{eqnarray}
\frac{1}{l^i_{total}} = \frac{1}{l^i_{abs}} + \frac{1}{l^i_{scatt}},
\label{totmfp}
\end{eqnarray}
with $i = n$ (neutron matter) or $q$ (quark matter). The radius and temperature dependence of the kick velocity together with the total neutrino mean free paths for quark and neutron matter as calculated in eq. (\ref{totmfp}) is plotted in Fig. \ref{vel_mfp}. Different neutron star velocities for a given initial temperature $T$ and a quark phase radius $R$ are shown. The stars are assumed to have a total mass of $1.4 M_\odot$. The dependencies of the kick velocity, the neutrino mean free paths and the quark phase radius on the quark and electron chemical potentials  $\mu_q$ and $\mu_e$ as well as $\alpha_s$ are shown in Figs. \ref{vel_mfp}-\ref{muq500}. For a temperature of 5 MeV the mean free paths in neutron matter and quark matter are in the range of only 100 m and therefore up to a factor 100 smaller than the quark phase radius.
For decreasing $\alpha_s$ eq. (\ref{cap_q}) predicts an increase of the heat capacity and consequently of the kick velocities (see eq. (\ref{velocity2})). The same holds for the neutrino mean free path for neutrino quark interactions as seen from eqs. (\ref{1overlabs}), (\ref{qfm}) and (\ref{li}), whereas the absorption and scattering of neutrinos in hadronic matter remain unchanged (see eq. (\ref{result3}) and eq. (\ref{result4})). A value of $\alpha_s \sim 0.3$ is normally expected for quark chemical potentials in the range of 1 GeV \citep{pdbook} . But even for such a low value we see from Fig. \ref{var_alpha} that the change of the neutrino mean free path in quark matter and the kick velocities are very small. In Fig. \ref{var_muq} we use two different quark chemical potentials of $\mu_q=300$ MeV and $\mu_q=500$ MeV. From eq. (\ref{cap_q}) and eq. (\ref{velocity2}) we expect again an increase of the kick velocity but also a decrease of neutrino mean free path for $\mu_q=500$ MeV in comparison to Fig. \ref{vel_mfp} where the quark chemical potential was set to 400 MeV. For $\mu_q=300$ MeV, the neutrino quark scattering and absorption mean free paths are larger but the kick velocities decrease. Consequently, we applied a combination of a small strong coupling constant $\alpha_s=0.3$ to increase the neutrino mean free path and a large quark chemical potential $\mu_q=500$ MeV to get higher kick velocities. The results are shown in Fig. \ref{0.3_500_10}. For quark phase radii of approximately 20 km and temperatures of $\sim$0.5 MeV a kick velocity of 100 km s$^{-1}$ is reachable. Of course such a large radius is not realistic for a quark star.   Consequently neither the variation of the strong coupling constant $\alpha_s$ nor the change of quark chemical potential will give a considerable improvement for the final kick velocities. As can be seen in Fig. \ref{0.3_500_T} a significant change in the neutrino-quark reaction rate can be achieved by lowering the electron chemical potential. However, the original mean free paths taken from \citet{Iwamoto81} were calculated assuming degenerate electrons, i.e. $\mu_e\gg$T. Consequently, for $T>\mu_e$ the electron chemical potential in eq. (\ref{result1}) is replaced by the temperature, as shown by \citet{Burrows80} for the neutrino emissivity in the direct quark Urca process. Electron-positron pair production due to the high temperature and their effects on the neutrino interaction rate have been ignored.\\
The neutrino mean free path in quark matter hits the 100 km s$^{-1}$ velocity curve for $R\sim $11 km and $T\sim$ 1.2 MeV. The 200 km s$^{-1}$ line is reached for $R\sim$15 km and $T\sim$ 1 MeV. In these cases the corresponding mean free paths in neutron matter are in the range of $\sim$ 6 km. Hence, in this simple approximation, which has to be verified in a future work, one can construct pulsar kicks with a velocity in the range of 100 km s$^{-1}$ and possibly 200 km s$^{-1}$. The magnetic field which is required in this case to polarise the electrons can be taken from Fig. \ref{pola} and is in the range of $\sim 10^{16}$ Gauss. The condition $\mu_e\ll T$ can be realised in a colour superconducting quark phase due to the presence of strange quarks (the CFL phase does not require any electrons for charge neutrality \citep{Rajagopal01}). Note, that in unpaired strange quark matter $\mu_e \simeq {m_s^2}/{4\mu_q}\sim 5-6$ MeV for $m_s =100$ MeV and $\mu_q = 400-500$ MeV. Strange stars have another feature which would boost the maximum possible kick velocity. The size of the hadronic phase for a hybrid star can vary between 1 and 3 km see e.g. \citet{Schertler1}. A strange star with M$_{ns}=1.5 M_\odot$ can have a nuclear crust of only 100 m - 500 m thickness \citep{Weber92}. Recently, it was proposed by \citet{Reddy} that the crust of strange stars can be composed of strange quark nuggets and electrons. The thickness of such a crust is calculated to be just 50 m. The neutrino mean free path is then similar to a bare strange star and one can ignore effects from the neutrino interactions, so that the final kick is maximal. 
\section{Effects of colour-superconductivity}
\label{effects_csc}
\begin{figure}
{\centering \includegraphics[width=0.34\textwidth, angle=270]{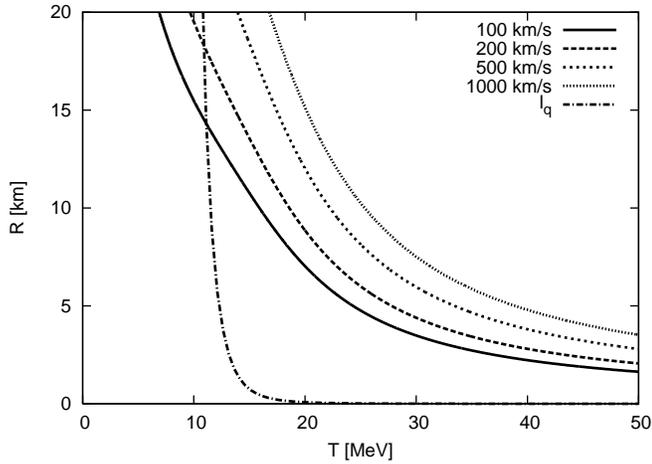} \par}
\caption{Kick velocities and neutrino mean free path in quark matter for the CFL quark phase with $\alpha_s=0.5$, $\mu_q=450$ MeV, $\mu_e=100$ MeV and $\Delta=100$ MeV.}
\label{mfp_gap}
\end{figure}

In the previous sections we saw that it is difficult to evade the problem of the high neutrino interaction rates in hot and dense matter by varying $\alpha_s, \mu_e$ or $\mu_q$. Consequently, we have to look for another mechanism. The neutrino mean free path is certainly effected by strong magnetic fields. In particular electrons are bound to Landau levels. The energy separation between two Landau levels is $\sqrt{2eB}$ which would be around $30$ MeV for $B=1\cdot 10^{17}$ Gauss and in the range of 10 MeV for $B=10^{16}$ Gauss. The quantised energy levels act as a gap and suppress the neutrino quark interaction at least in the plane perpendicular to $B$. Another suppression mechanism arises in quark matter if it is in a colour superconducting phase. If the temperature is lower than a critical temperature pairing of quarks increases the neutrino mean free paths. To suppress neutrino quark interactions in general all quarks have to be paired which is the case in the CFL phase. Here charge neutrality is provided by the equal amounts of down, up and strange quarks which means that the presence of electrons is not required and therefore the CFL phase has  $\mu_e=0$ \citep{Rajagopal01}. On the other hand, electrons can be present in the system via electron-positron pair production at finite temperature. In addition, a metallic CFL (mCFL) phase with nonzero $\mu_e$ can appear for $\beta$-stable matter for temperatures larger than $\sim$10 MeV \citep{Ruester04}. Large electron chemical potentials are found for $\mu_q \sim 400$ MeV \citep{Ruester}. Consequently, it would be interesting to study the neutrino emission from quark matter in the mCFL phase as the latter provides the quark core with high $\mu_e$ as well as suppressed quark interactions. Unfortunately, the pairing between quarks decreases the quark heat capacity by:
\begin{eqnarray}
c_q=9\left(1-\frac{2\alpha_s}{\pi}\right){\mu_q}^2Te^{-\Delta(T)/T},
\label{gap_hc}
\end{eqnarray}
where $\Delta$ is the gap energy \citep{Blaschke01}. For large values of $\Delta$, $c_q$ is lowered drastically. Consequently, the electron heat capacity 
\begin{eqnarray}
c_e=\frac{{\mu_e}^2T}{2}+\frac{7}{30}g\pi^2T^3,
\label{el_hc}
\end{eqnarray} 
ignored in section \ref{h_cap}, becomes important. Implementing both heat capacities in eq. (\ref{velocity1}) gives the following result for the kick velocity:
\begin{eqnarray}
v&=&\frac{2}{3}\pi R^3\frac{\chi}{M_{ns}}T^2\left(\frac{{\mu_e}^2}{2}+\frac{7\pi^2 T^2}{60}\right)^2
\nonumber\\
&+& \frac{2}{3}\pi R^3\frac{\chi}{M_{ns}}T^29\left(1-\frac{2\alpha_s}{\pi}\right){\mu_q}^2 e^{-\Delta/T}.
\label{new_vel}
\end{eqnarray}
As was shown by \citet{Ruester} and \citet{Blaschke2} large values for the gap $\Delta$ as well as for the electron chemical potential $\mu_e$ can arise in the colour superconducting quark core. Since we just want to study the behaviour of the neutrino mean free paths and the kick size in principal, we will use the following estimates: we choose the quark chemical potential to be 450 MeV (see  \citet{Ruester} and \citet{Blaschke2}), $\mu_e = 100$ MeV and $\Delta=100$ MeV. The kick velocity and the mean free paths are plotted in Fig. \ref{mfp_gap} for $\alpha_s=0.5$, $\mu_q=450$ MeV with temperatures up to 50 MeV. These values for $T$ are of course quite high, but well in the range of proto-neutron star evolution (see e.g. \citet{Lattimer04}).
Comparing Figs. \ref{vel_mfp} and \ref{mfp_gap} we see the consequence of the decreased quark heat capacity. For radii $R<10$ km velocities of 1000 km s$^{-1}$ can be reached for $T >$ 30 MeV. For increasing temperatures the suppression factor from the gap, $e^{-\Delta/T}$, diminishes and the contribution from quark matter becomes dominant again. As all neutrino-quark interactions are suppressed the neutrino mean free path is greatly enhanced for $T\ll\Delta$. However, the kick velocity is reduced at high temperatures so that just values for $v < 100$ km s$^{-1}$ can be reached for free streaming neutrinos and a realistic quark phase radii. As can be seen from Fig. \ref{mfp_gap} a quark phase radius of $\sim$14 km is required to accelerate the star to 100 km s$^{-1}$. Regarding the high temperatures as well as the high electron chemical potential discussed here the magnetic fields for fully polarised electrons are expected to be extremely high. For $T>$ 10 MeV and electron chemical potentials of $\mu_e>80$ MeV the magnetic fields must be larger than  $10^{18}$ G and therefore higher than the critical value of $\sim 1.3\cdot 10^{18}$ Gauss for stability \citep{Shap_Lai}. \citet{Reddy02} calculated neutrino mean free paths in the colour flavour locked phase to be larger than 10 km for a temperature of $T=5$ MeV. However, they found that the interaction with massless bosons which appear due to the breaking of baryon number conservation in the CFL phase leads to a mean free path which is similar or even shorter than the one for normal quark matter. For temperatures of around 30 MeV this means a very small value for the mean free path in the range of cm which would make the acceleration mechanism even in gapped quark matter infeasible. From eq. (\ref{vel_gen}) we know that a large kick velocity requires a sufficient energy release. Up till now the neutrino energy was given by the temperature. In the case of a phase transition from an unpaired to a gapped quark matter phase with the gap $\Delta$ an energy of $3\Delta^2 \mu_q^2/\pi^2$ would be released \citep{Alford01a}. If we assumed a neutrino emission of one neutrino per quark, the energy per neutrino would be $\Delta^2/\mu_q$. For the case $\Delta^2/\mu_q \sim 1$ MeV  the gap would be in the range of 20 MeV for $\mu_q \sim 450$ MeV. Consequently, for low temperatures the neutrino mean free path would be large as seen from eq. (\ref{li}) and eq. (\ref{1overlabs}). At the same time the energy release $3\Delta^2 \mu_q^2/\pi^2 \sim 3.2$ MeV/fm$^3$ would be sufficient to accelerate the quark star to 1000 km s$^{-1}$ if the quark phase radius was in the range of 8 km (see eq. (\ref{acc_mev})). In this connection it is also interesting to study a scenario which was proposed by \citet{Lavagno07}, where the authors discuss the transformation of a hadronic star into a quark star or a hybrid star. They find the formation of a convective layer in the case of conversion from ungapped (e.g. 2SC) to gapped quark (CFL) matter. An asymmetry in the fast transport of hot CFL quark matter to the surface of the hybrid or quark star could create large kicks.
\section{Summary and Outlook}
In this work we studied an acceleration mechanism for pulsars based on asymmetric neutrino emission from the direct quark Urca process in the interior of magnetised proto neutron stars. The anisotropy arises due to a strong magnetic field which forces the electrons into the lowest Landau level where their spin is polarised opposite to the magnetic field direction. To fully spin polarise the electrons for temperatures of 1-10 MeV we find the required magnetic field strength to be of the order $10^{16}$ - $10^{18}$ Gauss, depending on the electron chemical potential. The pulsar kick velocity depends on the released energy in the quark phase and stems in our case from the heat reservoir. For fully spin polarised electrons we find that for an initial temperature of $T> 5$ MeV a kick velocity of 1000 km s$^{-1}$ is possible for a radius of $R<$10 km and a mass of 1.4 M$_\odot$ suitable for a strange star. The required magnetic fields have to be in the range of $B\sim 10^{17}$ G to achieve full polarisation for electron chemical potentials of $\mu_e <30$ MeV. However, as shown by \citet{Vilenkin} high neutrino quark interaction rates will wash out the beamed flux of neutrinos necessary for the acceleration mechanism. For typical properties of quark matter in the interior of a quark star the neutrino mean free paths are in the range of 100 m to 800 m \citep{Iwamoto81} which is, at least for the quark matter core, too small. The best results were found for $\alpha_s=0.5$, $\mu_q=400$ MeV and $\mu_e\sim T$ giving a velocity of 100 km s$^{-1}$ for quark phase radii of $R\sim 11$ km and a temperature in the range of $T \sim 1$ MeV. A kick velocity of 200 km s$^{-1}$ was reached for $R\sim 15$ km and $T \sim 1.2$ MeV. For both cases the required magnetic field is in the range of $\sim$10$^{16}$ Gauss. To suppress the neutrino quark interaction and to accommodate a small electron chemical potentials we investigated in a first study effects from colour superconducting quark matter in the CFL phase. The neutrino mean free path is enlarged by a factor $\exp(-\Delta/T)$ where $\Delta$ is the pairing energy of the quarks. Unfortunately, the quark heat capacity also decreases exponentially with  $\Delta/T$ so that the largest reachable kick velocity is again 100 km s$^{-1}$ for a gap of $\Delta= 100$ MeV, $\mu_e =100$ MeV, a quark chemical potential $\mu_q=450$ MeV and a quark phase radius of $R=$14 km. In summary, the mechanism discussed here can produce neutron star velocities of 1000 km s$^{-1}$ very easily if neutrino quark interactions are ignorable. However, the small neutrino mean free paths seems to make it impossible to reach velocities higher than 100 km s$^{-1}$. For the neutrino mean free path to be large small temperatures are required. At the same time the energy release should be large to accelerate the compact star to high velocities. A solution might be a phase transition from unpaired to gapped quark matter as discussed by e.g. \citet{Lavagno07}, where the energy release is about $3 \Delta^2 \mu_q^2/\pi^2$. The energy carried off by neutrinos would be $\sim \Delta^2/\mu_q$ per neutrino which is in the range of 1 MeV for $\mu_q \sim 450$ MeV and $\Delta \sim 20$ MeV and therefore very small. The total released energy is very large and can accelerate the compact star to 1000 km s$^{-1}$ if the radius of the converted quark phase is around 8 km. In this connection it is interesting to study the phenomenon discussed by \citet{Lavagno07}. The authors find the creation of a convective layer at the conversion zone which could transport the neutrino emitting gapped quark matter to the surface of the star and lead to a pulsar kick. In a future work we plan to add effects from quark pairing in the CFL phase as well as magnetic field effects for the heat capacities and particle energy densities. The energy release in neutrinos and therefore the kick velocities as well as the neutrino mean free path in the neutron star's interior will be studied in more detail.
\begin{acknowledgements}I. Sagert is supported by the Helmholtz Research School for Quark Matter studies in Heavy Ion Collisions.\end{acknowledgements}
\bibliography{all}
\end{document}